\documentclass[12pt]{article}
 \pdfoutput=1

\usepackage{amsmath,amssymb,amsfonts}
\usepackage{graphicx}
\usepackage{color}
\usepackage{amsmath}
\usepackage{amsfonts}
\usepackage{amssymb}
\usepackage{float}
\usepackage{cancel}
\usepackage{hyperref}

\numberwithin{equation}{section}

\newcommand{\beq}{\begin{eqnarray}}
\newcommand{\eeq}{\end{eqnarray}}

\newcommand{\centeron}[2]{{\setbox0=\hbox{#1}\setbox1=\hbox{#2}\ifdim
\wd1>\wd0\kern.5\wd1\kern-.5\wd0\fi \copy0
\kern-.5\wd0\kern-.5\wd1\copy1\ifdim\wd0>\wd1
                                  \kern.5\wd0\kern-.5\wd1\fi}}
\newcommand{\ltap}{\>\centeron{\raise.35ex\hbox{$<$}}
                          {\lower.65ex\hbox{$\sim$}}\>}
\newcommand{\gtap}{\>\centeron{\raise.35ex\hbox{$>$}}
                          {\lower.65ex\hbox{$\sim$}}\>}

\newcommand\ZZ{\hbox{\zfont Z\kern-.4emZ}}
\font\zfont = cmss10 

\newcommand{\fref}[1]{Fig.\ \ref{f.#1}}

\newcommand{\sref}[1]{Section \ref{s.#1}}
\newcommand{\cref}[1]{Chapter \ref{c.#1}}

\def\nn{\nonumber \\}

\def\beq{\begin{equation}}
\def\eeq{\end{equation}}
\newcommand{\ba}{\begin{array}}
\newcommand{\ea}{\end{array}}
\newcommand{\bea}{\begin{eqnarray}}
\newcommand{\eea}{\end{eqnarray} }
\newcommand{\bal}{\begin{align}}
\newcommand{\eal}{\end{align}}
\def\bi{\begin{itemize}}
\def\ei{\end{itemize}}
\def\ben{\begin{enumerate}}
\def\een{\end{enumerate}}
\def\beq{\begin{equation}}
\def\eeq{\end{equation}}
\def\bc{\begin{center}}
\def\ec{\end{center}}
\def\bt{\begin{table}}
\def\et{\end{table}}
\def\btb{\begin{tabular}}
\def\etb{\end{tabular}}

\def\cl{{\mathcal L}}

\def\gev{\, {\rm GeV}}

\def\mass2{mass${}^2$}

\def\ra{\rangle}
\def\la{\langle}

\def\pa{\partial}

\def\simlt{\stackrel{<}{{}_\sim}}

\newcommand{\ti}{\tilde}
\def\hc{{\rm h.c.}}

\def\eps{\epsilon}

\newcommand{\zd}{\gamma_d}
\newcommand{\zdmunu}{\gamma_d^{\mu \nu}}
\newcommand{\fzd}{\tilde \gamma_d}
\newcommand{\mhid}{m_{\gamma_d}}
\newcommand{\mhhid}{m_{h_{d}}}
\newcommand{\Gd}{G_{d}}
 \newcommand{\nd}{\tilde n_{d}}
 \newcommand{\hd}{h_{d}}
 \newcommand{\thd}{\tilde h_{d}}
 \newcommand{\Hd}{H_{d}}
 \newcommand{\Ad}{A_{d}}
 \newcommand{\gd}{g_{d}}

 \newcommand {\unit} [1] {\; \mathrm {#1}}
 \newcommand{\muhid}{\mu_{d}}
 \newcommand{\Mhid}{M_{d}}
 \newcommand{\betahid}{\beta_{d}}
 \newcommand{\gammahid}{\gamma_d}
 \newcommand{\Bmuhid}{B\mu_{d}}

  \newcommand{\mE}{\slash \negthickspace \negthickspace E}

\begin{document}
\begin{titlepage}
\vskip1.5cm
\vskip1.cm
\begin{center}
{\huge \bf Hidden Higgs Decaying to \\ Lepton Jets}
\vspace*{0.1cm}
\end{center}
\vskip0.2cm

\begin{center}
{\bf  Adam Falkowski$^{a}$, Joshua T.~Ruderman$^{b}$, Tomer Volansky$^{c}$ and Jure Zupan$^{d}$}

\end{center}
\vskip 8pt

\begin{center}
{\it $^{a}$ NHETC and Department of Physics and Astronomy, 
\\ Rutgers University, Piscataway, NJ 08854, USA}
\\
{\it $^{b}$ Department of Physics, Princeton University, 
Princeton, 
NJ 08544}
\\
{\it $^{c}$ School of Natural Sciences, Institute for Advanced Study, 
Princeton, NJ 08540}
\\
{\it $^{d}$ Faculty of Mathematics and Physics, University of Ljubljana 
\\ Jadranska 19, 1000 Ljubljana, Slovenia}

\vspace*{0.3cm}

\end{center}

\vglue 0.3truecm

\begin{abstract}
  \vskip 3pt \noindent

  The Higgs and some of the Standard Model superpartners may have been
  copiously produced at LEP and the Tevatron without being detected.
  We study a novel scenario of this type in which the Higgs decays
  predominantly  into a light hidden sector either directly or
  through light SUSY states.  Subsequent cascades increase the
  multiplicity of hidden sector particles which, after decaying back
  into the Standard Model, appear in the detector as clusters of
  collimated leptons known as lepton jets.  We identify the relevant
  collider observables that characterize this scenario, and study a
  wide range of LEP and Tevatron searches to recover the viable
  regions in the space of observables.  We find that the Higgs
  decaying to lepton jets can be hidden when the event topology
  mimics that of hadronic backgrounds.  Thus, as many as $10^4$
  leptonic Higgs and SUSY decays may be hiding in the LEP and
  Tevatron data.  We present benchmark models with a 100 GeV Higgs
  that are consistent with all available collider constraints.  We end
  with a short discussion of strategies for dedicated searches at LEP,
  the Tevatron and the LHC, that allow for a discovery of the Higgs
  or SUSY particles decaying to lepton jets.

\end{abstract}

\end{titlepage}

\newpage

\tableofcontents

\section{Introduction}

Within the Standard Model (SM) the mass of the Higgs boson is
constrained by the LEP experiments to be larger than 114.4 GeV \cite{Barate:2003sz}.  This
limit, however, varies in a model-dependent manner.  For example, the
limit can be relaxed if the production cross-section of the Higgs
particle is suppressed.  A more interesting possibility is that the
{\em Higgs is hidden: it has been copiously produced at LEP and
  the Tevatron but has evaded detection due to non-standard decays}.
There are at least two hints that the Higgs might be lighter than the
naive LEP limit.  On the experimental side, the best fit to
electroweak precision observables corresponds to a Higgs mass of 80
GeV \cite{gfitter}.
On the theoretical side, typical supersymmetric models require
fine-tuning to accommodate a Higgs boson substantially heavier than
$m_Z$.

The naive LEP limit on the Higgs mass is based on studies of the
associated  production of the Higgs and $Z$-boson, with
the Higgs decaying to a pair of $b$ quarks (as in the SM).  
In theories beyond the SM, the Higgs decay pattern
can be greatly modified~\cite{O'Connell:2006wi}. 
Then, as long as the $h\to b\bar b$ branching ratio is below $\lesssim 20\%$, the standard Higgs search
strategies and mass limits may not directly apply~\cite{Lep_4b}.
The LEP collaborations have put considerable effort into constraining
non-standard Higgs decays into invisible
particles or into final states
with two SM particles.  However, Higgs decays into
higher multiplicity final states have not been systematically searched for.
Consequently, it is conceivable that the
Higgs decaying in a non-standard way may have been missed at
colliders, if it decays into a final state that has not been targeted.  In reality, it is not necessarily easy for a light Higgs to remain hidden, because the collective body of LEP searches constrain many different final state topologies.  Even if a specific Higgs decay mode has not been searched for directly, it may still be captured by a multitude of LEP searches. 

There are a few proposals for a hidden Higgs in the literature (for a review see~\cite{Chang:2008cw,Dermisek:2009si}).
The most studied scenario (and the first one to point out the improved naturalness with a hidden Higgs) is in the context of the NMSSM~\cite{DG1}.  
Additional examples include 
more general singlet extensions of the MSSM \cite{CFW}, supersymmetric little Higgs models~\cite{buried}, the R-parity violating MSSM~\cite{linda}, and the CP-violating MSSM~\cite{cpv}. 
In all of these scenarios the Higgs decays into a pair of light non-SM particles, e.g.
pseudoscalar singlets or neutralinos, which then decay
into two or more visible SM particles.  
Recently, the hidden Higgs scenario has triggered enough interest in
the community to prompt the revisiting of LEP data in search of certain
4-body Higgs decay topologies.  
The $h\to 4b$~\cite{Lep_4b} and $h\to 4\tau$~\cite{kyle} possibilities are now excluded for Higgs masses $\lesssim 110$ GeV\@.  
Other scenarios, however, are constrained only by the model-independent
Higgs search of OPAL~\cite{Opal_general}, requiring that the Higgs mass be above $\sim 82$ GeV. 

The hidden Higgs idea is most naturally expressed in the context of supersymmetry (SUSY) where the existence of a light Higgs,
$m_h \simeq m_Z$, ameliorates the little hierarchy problem.  In this
case it is natural to wonder whether {\em SUSY could also be hidden}: some of the SM
superpartners may have been copiously produced at colliders and
evaded detection due to non-standard decays.  
In this paper, we realize hidden Higgs and hidden SUSY in a
supersymmetric model with a light hidden sector.
The lightest `visible' superpartner (LVSP -- the equivalent of the LSP in the MSSM) is allowed to
cascade into the hidden sector, typically producing visible particles
in the process.  
The sensitivity of standard SUSY searches can then be greatly diminished when the LVSP decays partly into visible energy.

We consider three distinct scenarios where the lightest MSSM Higgs
boson decays dominantly into the hidden sector.
In the {\em singlet channel} scenario  the Higgs 
decays to the hidden sector through direct couplings.
In the other two scenarios the Higgs first decays to a pair of LVSPs which, 
having no visible decay channels open, decay into the hidden sector.  
The MSSM contains two types of electrically neutral and colorless superpartners which leads to  the 
{\em neutralino channel} and the {\em sneutrino channel} scenarios.  
In all of the models we consider, hidden sector cascades produce a
large multiplicity of boosted hidden sector particles.  Some of these hidden sector particles decay to leptons, and the final state of the
Higgs decay is therefore characterized by several groups of collimated leptons plus missing
energy.   The name {\em lepton jet} has been coined for these
spectacular objects~\cite{nn,itay, baumgart}.

The striking phenomenology of light hidden sectors has been studied in the past~\cite{kathryn}.  More recently, the existence of such sectors was motivated by the observed astrophysical anomalies~\cite{pamela,Collaboration:2008aaa,Abdo:2009zk}.   Indeed, the observed excesses in the positron and electron cosmic ray fluxes 
may be signatures of dark matter annihilations~\cite{tracy,MPSV,Pospelov:2008jd,Rothstein:2009pm,Cheung:2009qd,Chun:2008by} or
decays~\cite{MPSV,josh} into a light hidden sector which is weakly
coupled to the visible sector.
In this paper we do not address the aforementioned anomalies and concentrate instead, on the collider  signatures of such hidden sectors. 

 \begin{figure}[t]
\bc
\vspace{-0.5 cm}
\includegraphics[width=0.6\textwidth]{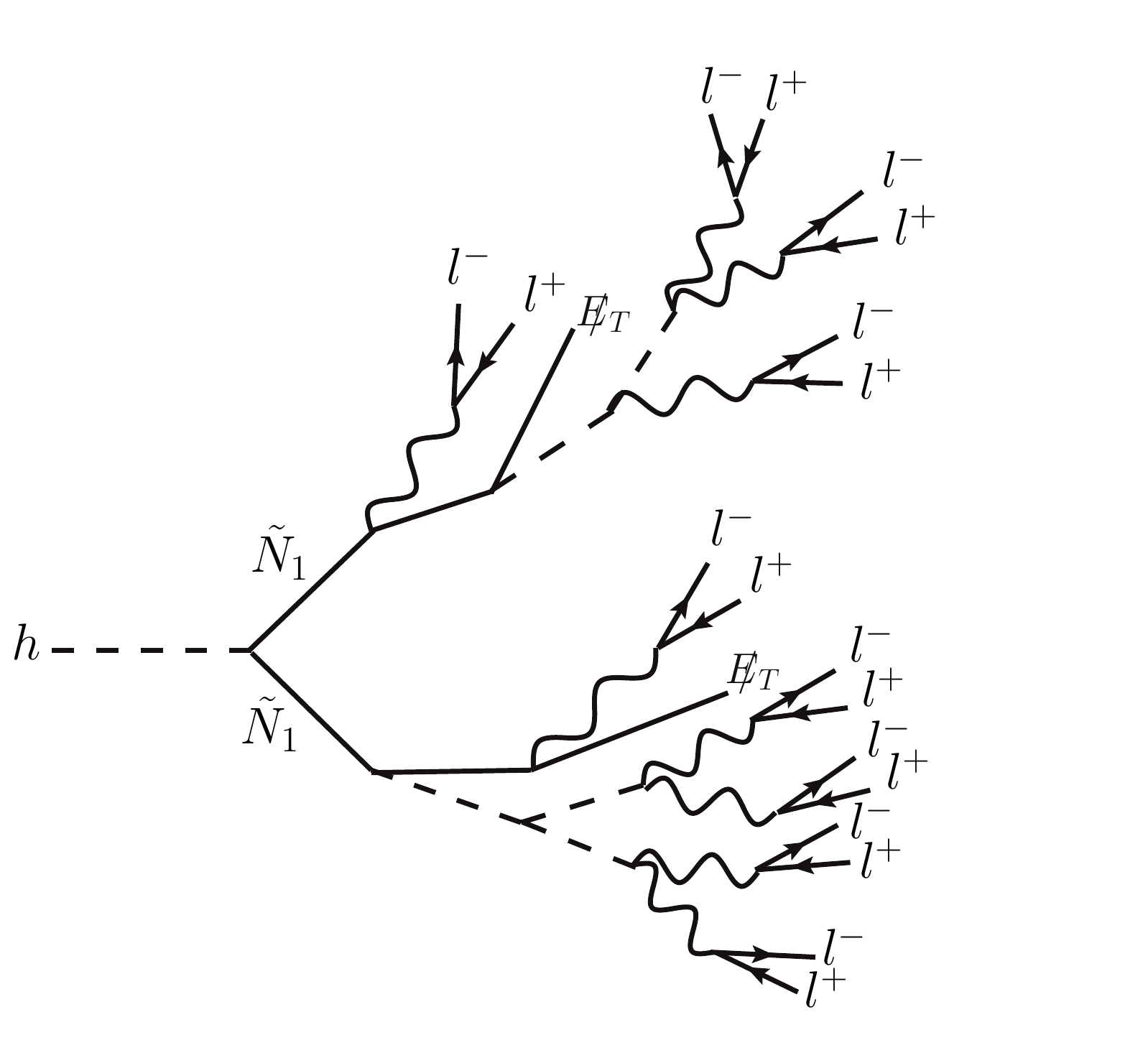}
\vspace{-0.7 cm}
\ec
\caption{\small
An example of a Higgs decay to lepton jets, through the neutralino production portal of Section \ref{sec:neutralino-channel}.  The hidden sector cascades can lead to many leptons per Higgs decay, in this case 18.  This example uses the particle content and vertices of the minimal $U(1)_d$ hidden sector described in section \ref{sec:minimal-model}. A larger hidden sector can lead to even larger multiplicities.  If the neutralinos are heavy enough to be produced close to rest, their decay products will be well-separated, and the leptons will partition into 4 distinct lepton jets.  Alternatively, if the neutralinos are light and boosted, the event will consist of two groups of collimated leptons, {\it neutralino jets}.}
\label{f.shd}
\end{figure}

As a simple example, we consider a hidden sector with $U(1)_d$ gauge symmetry broken at the GeV scale.  $U(1)_d$ couples to the visible sector through kinetic mixing with hypercharge, implying that (i) the hidden photon can decay to the light SM fermions, and (ii) the LVSP can decay to the hidden sector.   Consequently, once the Higgs decays, it initiates a hidden sector cascade, producing in addition to the true LSP, many hidden photons and scalars which decay to highly boosted lepton jets. An example of such a Higgs decay is shown in Fig.~\ref{f.shd}. 
To demonstrate that a light Higgs can be accommodated in the above scenario, we simulate Higgs decays to lepton jets and determine the sensitivity of a wide range of LEP and Tevatron searches.   We consider the experimental observables that are relevant for Higgs decays into lepton jets,  and identify the viable region in the space of these observables. 
This procedure allows us to write benchmark models for the neutralino and singlet channels in which the Higgs as light as 100 GeV is allowed by all existing searches. (The sneutrino channel is harder to accommodate with the specific hidden sector we consider.) 
One should stress that the final states in these models are so spectacular that a {\em dedicated} analysis at LEP or the Tevatron could quickly
discover the Higgs signal, or place a far more stringent bound on the Higgs boson mass.

Several previous studies have considered Higgs decays to light hidden sectors.  Ref.~\cite{Strassler:2006ri} considers decays that produce very displaced vertices.  While it is conceivable that such a scenario can accommodate a light Higgs, this possibility was not explored by the authors of~\cite{Strassler:2006ri} and we do not consider it here.  Ref.~\cite{james} discusses Higgs decays to two hidden sector photons which subsequently decay to four SM leptons~\cite{james}.  The authors do not attempt to hide a light Higgs in this scenario (but see comments in \cite{Dermisek:2009si}).
We take a complimentary approach by considering different models that yield a larger variety of final state topologies and multiplicities than these previous works, and in doing so, we identify scenarios that allow for a light Higgs boson with prompt decays. 
  
This paper is organized as follows.  In~\sref{hs} we review the
concept of a GeV-scale hidden sector that communicates with the SM through kinetic mixing, and introduce a minimal phenomenological model with a $U(1)_d$ gauge symmetry. 
In~\sref{p} we 
describe the channels via which the Higgs can 
decay into
the hidden sector. 
In~\sref{cp}
we discuss the collider phenomenology of Higgs decays to lepton jets by defining the experimental observables that characterize this scenario.  In Section~\ref{sec:crazy} we briefly explain why a light Higgs decaying into lepton jets is not obviously excluded by LEP and Tevatron searches, as one may naively suspect. 
We then review the relevant LEP and Tevatron searches
in~\sref{ec}.  In Section~\ref{sec:benchmark-models}, we discuss how these searches constrain the experimental observables, and construct benchmark models with a 100 GeV Higgs that satisfy these constraints.
 In~\sref{ss} we
discuss search strategies at LEP and the Tevatron that can differentiate lepton jets from QCD jets and allow for discovery of a light Higgs decaying to lepton jets.  We conclude in~\sref{c}.  We describe our hidden sector notations in Appendix~\ref{a.abp} and list benchmark signal efficiencies for LEP and Tevatron searches in Appendix~\ref{a.e}.

\section{The Hidden Sector}
\label{s.hs}

We begin by reviewing the framework in which we will later embed a hidden
Higgs.  In Section \ref{s.ohs}, we discuss portals that can connect a light hidden sector to the visible sector.  These portals take the form of operators composed of both hidden sector fields and SM fields.  We focus on the vector portal -- kinetic mixing between hidden sector gauge fields and SM gauge fields.  Then, in Section \ref{sec:minimal-model}, we specialize to a minimal phenomenological hidden sector, with $U(1)_d$ gauge symmetry, and discuss the interactions among hidden sector fields.  These hidden sector interactions will be important for the collider phenomenology discussed later in the paper.

\subsection{Portals}
\label{s.ohs}

To trigger non-conventional Higgs decays, we study a hidden sector
with a gauge group $\Gd$ broken at the GeV scale and weakly coupled to
the SM\@.  Here and below we work in the supersymmetric framework, which
allows one to stabilize both the weak and GeV scales.  For
simplicity, we will focus on $\Gd=U(1)_d$. 
We will see that this simple case is rich enough to allow for Higgs decays with tens of lepton tracks.
Non-Abelian models generalize the structure and provide a simple way
to further soften and  increase the multiplicity of the produced leptons.  We return to this
scenario in Section~\ref{subsec:benchmarks}.

\begin{figure}[t]
\hbox{
\includegraphics[width=0.35\textwidth]{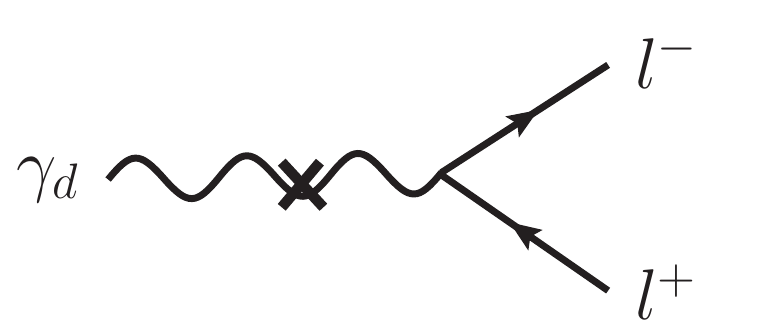}
\qquad 
\qquad 
\includegraphics[width=0.55\textwidth]{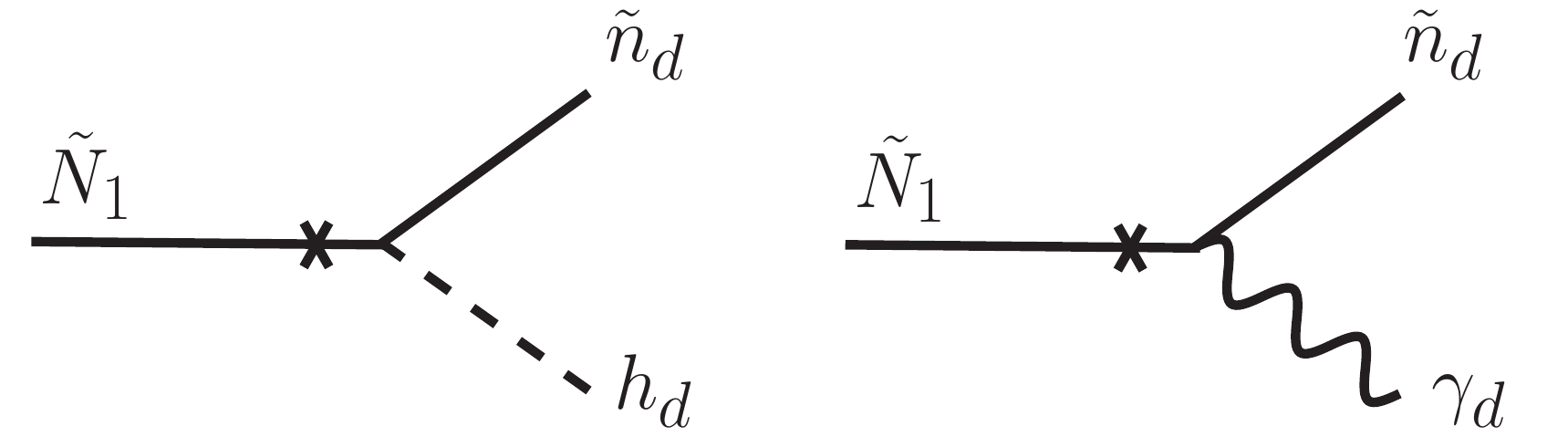}
}
\caption{\small  \setlength{\baselineskip}
{0.9\baselineskip}Interactions that follow from kinetic mixing between the hidden photon and hypercharge.  The hidden photon couples to the electromagnetic current, including lepton pairs, as in the diagram on the {\it left}. The cross on the diagram indicates the $\epsilon$ suppression.  The {\it right} two diagrams show possible decays of the SM bino to the hidden sector, which follow from gaugino kinetic mixing.  The SM LSP is no longer stable, and all SUSY cascades can end in the hidden sector.} 
\label{f.KMcoup}
\end{figure}

The hidden sector may couple to the visible sector through various
portals (for a useful discussion see~\cite{Batell:2009di}).  Here we
concentrate on the so-called {\em vector portal} which has been studied
extensively~\cite{Holdom:1985ag,Pospelov:2007mp,Batell:2009di,tracy,baumgart,Batell:2009yf,Essig:2009nc,Cheung:2009qd, MPZ}.
It is straightforward to extend this scenario to other portals.  The
communication of the hidden sector with the MSSM is through
kinetic mixing of the hidden photon, $\zd$, of mass $\mhid$ and the
hypercharge field $B_\mu$,
\begin{equation}
\label{e.km} 
\cl_{mix} =  {1 \over 2} \eps \, \zdmunu B_{\mu\nu} 
 = {1 \over 2} \eps \, \zdmunu \left (\cos\theta_W A_{\mu\nu} - \sin\theta_W Z_{\mu\nu} \right) \,.
\end{equation}
Here $\zdmunu \, (B_{\mu\nu})$ is the field strength of $\zd \, (B_\mu)$
and $\theta_W$ is the Weinberg angle.  The mixing parameter, $\eps$,
is assumed to be small, $\epsilon \lesssim 10^{-3}$.  The
mixing with the photon can be removed by a shift of the photon
field,
\begin{equation}
\label{eq:1}
 A_\mu \to A_\mu + \eps \cos 
\theta_W \zd  \, .
 \end{equation}
 As a consequence, the hidden photon couples to all electrically
 charged particles with the strength $\eps \, e \cos \theta_W$.  The
 smallness of $\eps$ implies millicharged couplings, consistent with all
 current bounds \cite{Batell:2009yf,Essig:2009nc}.  The main
 significance of the above mixing is to trigger the decay of the
 hidden photon, $\zd$, to kinematically accessible leptons and hadrons. 
 This is illustrated with the left diagram of Fig.~\ref{f.KMcoup}.  
 Decays of the hidden photon to electrically neutral particles that couple to the $Z$, such as neutrinos, are suppressed by  $\mhid^2/m_Z^2$ and will not play an important role.
 Similarly, we can ignore the mixing between the hidden and visible Higgses
 through the D-terms.

 Upon supersymmetrizing Eq.~(\ref{e.km}),   the hidden gaugino and visible
 bino (and therefore neutralinos) mix.  One finds the kinetic mixing
 terms
 \begin{equation}
- i \eps \, \fzd^\dagger \, \bar  \sigma^\mu \pa_\mu \ti B - i \eps \, \ti
B^\dagger \, \bar  \sigma^\mu \pa_\mu \fzd \,.
\end{equation}
Much as before, it is convenient to shift the hidden bino 
\begin{equation}
\ti \zd \to \fzd + \eps \, \ti B\,,
 \end{equation}
 which removes the kinetic mixing, while keeping the mass matrix
 diagonal to order $\mhid/m_Z$.  Consequently, hidden fields charged
 under $U(1)_d$ interact with the visible neutralinos with
 $\epsilon$-suppressed couplings.  
 In particular, all hidden sector scalars $\hd^i$, with charges $q_i$, couple to the visible bino,
 \begin{eqnarray}
   \label{eq:9}
    -\eps \, \gd \ti B\sum_i q_i \, \hd^{i\dagger} \, \thd^i \,.
 \end{eqnarray}
 Thus the visible neutralino may decay to the hidden neutralinos and
 either a Higgs or the longitudinal mode of the hidden photon, as in the right two diagrams of
 Fig.~\ref{f.KMcoup}.  In addition to the above, the small
 off-diagonal terms in the mass matrix induce $\mhid/m_Z$-suppressed
 couplings of the form
 \begin{eqnarray}
   \label{eq:10}
  -\epsilon \, g^\prime\, \frac{\mhid}{m_Z} \, \fzd \sum_i Y_i f^i\tilde f^i\,,
 \end{eqnarray}
 where $f^i$($\tilde f^i$) are MSSM (s)fermions, $Y_i$ are the
 corresponding hypercharges, and $g^\prime$ is the hypercharge
 gauge coupling.  This coupling will play a role in the sneutrino
 decay channel discussed in Section~\ref{sec:sneutrino-channel} below.

Finally, the hidden sector may have additional couplings to the
visible sector.  For instance, couplings of singlets in the
superpotential,
\begin{eqnarray}
  \label{eq:2}
 W \supset  S \, (y \, \chi\bar\chi + \lambda \, H_u H_d)\,,
\end{eqnarray}
can lead to the {\em higgs portal}, 
$H_u H_d \chi^* \bar \chi^* + \mathrm{c.c.}$, where $\chi$ and $\bar \chi$ are charged under the hidden sector.  If 
$\chi$ and/or  $\bar \chi$ get VEVs, this operator can trigger mixing between the MSSM Higgses and hidden particles.  If 
$\chi$, $\bar \chi$ are relatively heavy the presence of the mixing
does not change the decay branching fractions of the low-lying hidden
states.  
The Higgs portal can, however, lead to Higgs decays into hidden sector particles.  From now on we will refer to this mechanism as the {\em singlet channel}, which we return to in Section \ref{sec:singlet-channel}.

\subsection{A Minimal Model}
 \label{sec:minimal-model}

We now introduce the field content of a minimal hidden sector that leads to Higgs decays to lepton jets. 
In order to break the $U(1)_d$ gauge group, two Higgs chiral superfields, $h_{1,2}$ with
charges $\pm 1$, are added. Generally, both Higgs fields obtain VEVs.   This hidden sector setup was also studied in~\cite{itay} 
(and we give further details in Appendix~\ref{a.abp}).
The hidden spectrum
contains
\begin{itemize}
\item One massive photon $\zd$,  
\item Three hidden neutralinos $\nd^i$, which  are mixtures of the
  hidden gaugino and Higgsinos,
\item Three hidden scalars $\hd^i$: taking the hidden sector vacuum to preserve CP, there are two CP-even scalars, $\hd,\Hd$, and one
  CP-odd scalar, $\Ad$.
 \end{itemize}
 All these particles have masses which, for definiteness, are assumed
 to lie in the 100~MeV to few~GeV  ballpark.  Supersymmetry is softly broken
 in the hidden sector, and we assume completely general soft and $\mu$
 terms.  This gives sufficient freedom to organize the masses in the
 hidden sector into various patterns leading to different types of
 cascades. The interactions within the hidden sector are fully dictated
 by gauge symmetry and supersymmetry.  The neutralinos interact via,
 \begin{eqnarray}
   \label{eq:3}
   \nd^i \, \nd^j \, \hd^k \,,\qquad \qquad \nd^{i\dagger} \, \sigma_\mu \nd^j \, \zd^\mu, 
 \end{eqnarray}
 where the couplings are fixed by the
 hidden gauge couplings and mixing angles.  Through these vertices the
 hidden neutralinos can cascade down to the lightest one (which we
 assume to be the true LSP\footnote{If visible-sector SUSY is broken by gauge mediation, there will also be a light gravitino.  Hidden fermions decay to the gravitino well outside the detector \cite{baumgart}, so here we can neglect the gravitino.}, emitting hidden scalars and photons).
 Whenever it is kinematically available, the scalars can also decay
 through the vertices
 \begin{eqnarray}
   \label{eq:4}
    \hd^i \, \hd^j \, \hd^k \,,\qquad \qquad \hd^i \, \zd^\mu \, (\zd)_\mu ,
 \end{eqnarray}
that originate from the D-term and from the scalar kinetic
 terms.  
 
 Thus, depending on the mass spectrum, the cascades may lead
 to a large multiplicity of hidden particles in each event. 
 The mass spectrum controls, for example, the
 typical length of the hidden cascade, the multiplicity of visible
 final states, and the amount of missing energy.  The minimal model can be deformed by considering a non-Abelian hidden
 gauge group or by adding more chiral multiplets, both of which
 increase the number of scalar and fermionic eigenstates and can 
 lengthen the hidden cascades, thereby producing a larger final state
 multiplicity.  We study an example with such a modification in Section~\ref{subsec:benchmarks}.

\section{Higgs Decays to the Hidden Sector}
\label{s.p}

The MSSM by itself allows for a rich variety of Higgs decay modes,
depending on the visible spectrum.  In particular, if the LVSP
is sufficiently light, the Higgs can decay to it with a sizable
branching fraction, much larger than to
the SM channels like $b\bar b$
or $\tau^+\tau^-$.  If the LVSP is stable, such a scenario is strongly constrained
by invisible Higgs  searches at LEP~\cite{Opal_invisible}.  The
mixing of the hidden sector with the MSSM, however, makes  
the LVSP 
unstable. 
Then the Higgs can decay 
predominantly into complicated
high-multiplicity final states.  Such a possibility has not been
experimentally studied at LEP or the Tevatron, therefore a priori a Higgs
boson can be lighter than the naive LEP limit of 114.4 GeV\@.
Below we identify three possible channels through which the lightest
CP-even Higgs of the MSSM can decay into the hidden sector.

\subsection{Neutralino Channel}
\label{sec:neutralino-channel}

\begin{figure}[t]
\bc
 \includegraphics[width=0.6\textwidth]{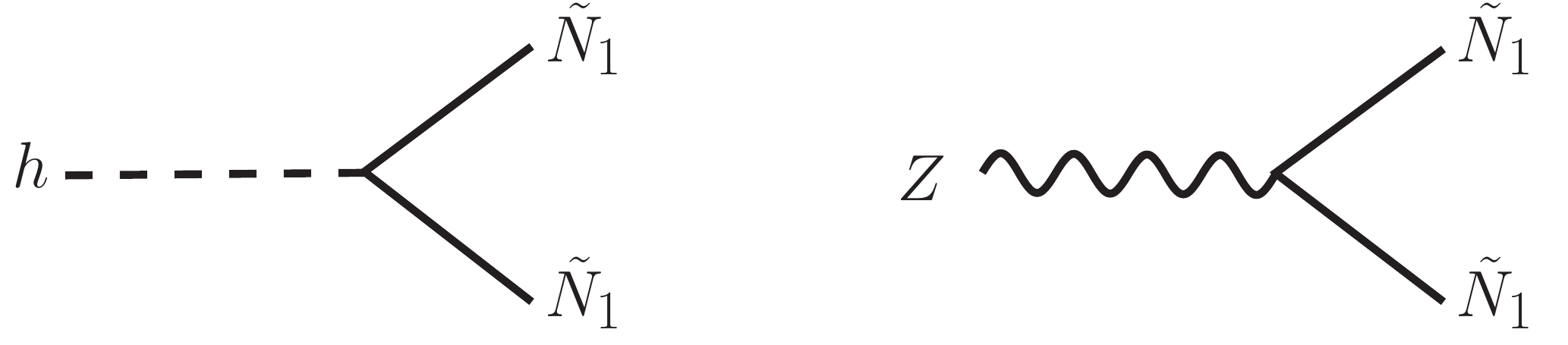}
\ec
\caption{\small \setlength{\baselineskip}
{.9\baselineskip}  Higgs and $Z$ decays to neutralinos.  The neutralinos then decay into the hidden sector as in figure \ref{f.KMcoup}.   We consider the region of the MSSM parameter space where the Higgs dominantly decays to neutralinos while the charginos are above the LEP bound of $\sim100$ GeV\@.  In this region, we find $m_{\tilde N_1} < m_Z/2$, such that the $Z$ can also decay to neutralinos, as in the {\it right} diagram.  This is consistent with LEP-1 constraints when $\mathrm{BR} \, (Z \rightarrow  2 \,\tilde N_1) \lesssim 10^{-3}$.}
\label{f.np}
\end{figure}
 
In principle, there is no model-independent bound on the mass of the
LVSP neutralino (this is because the bino is an MSSM singlet).   The bounds on the Higgs
boson mass can then be significantly relaxed, if the
LVSP neutralino is sufficiently light, such that the Higgs branching fraction into it is
$\gtrsim75\%$, while a sizable fraction of the neutralino energy
comes back in the form of visible SM states.   A similar scenario with a
light Hidden Higgs was considered  in Ref.~\cite{linda} within the R-parity violating MSSM,
where the neutralino decays into three SM quarks, leading to the
Higgs-to-6-jets signature.  Higgs decays to neutralinos have also been considered in the framework of gauge mediation in~\cite{Mason:2009qh}.  Here we revisit the
neutralino channel in the context of the Higgs decaying to lepton jets.

The coupling of the lightest MSSM Higgs boson to the lightest
neutralino arises from the Higgs-Higgino-Bino/Wino couplings and takes
the form,
\begin{eqnarray}
  \label{eq:5}
   g_{h11} h \ti N_1 \ti N_1 + \hc \,, \qquad g_{h11} = {1 \over 2}
   \left ( g c_{W} - g' c_{B} \right ) \left(c_\alpha c_{U} + s_\alpha c_{D} \right ) \,.
\end{eqnarray}
We parametrized the
embedding of the lightest CP-even Higgs boson into the Higgs doublets
as $H_u^0 = (s_\beta v + c_\alpha h + \dots)/\sqrt{2}$, $H_d^0 =
(c_\beta v - s_\alpha h + \dots)/\sqrt{2}$.   The angles $c_i$ describe
the composition of the lightest neutralino in terms of the original
gauginos and Higgsinos: $\ti N_1 = c_B \ti B + c_W \ti W^3 + c_U \ti
H_u^0 + c_D \ti H_d^0$.  The Higgs partial decay width is,
\begin{eqnarray}
  \label{eq:6}
  \Gamma(h\to \ti N_1 \ti N_1) = {g_{h11}^2 m_h \over 4 \pi} \left (1 -
  4 {\ti m_{N1}^2 \over m_h^2} \right )^{3/2} \,.
\end{eqnarray}
This should be
compared with the decay width into a pair of $b$-quarks,
\begin{eqnarray}
  \label{eq:7}
   \Gamma
(h\to b \bar b) = c_{\rm QCD} {3 \over 8 \pi} y_{hbb}^2 \left (1 - 4
  {m_b^2 \over m_h^2} \right )^{3/2} \,,
\end{eqnarray}
where $y_{hbb} = c_\alpha m_b/c_\beta v$ and $c_{\rm QCD}$ is a fudge
factor that captures higher-order QCD effects.  The latter are
numerically relevant; for example, $c_{\rm QCD} \approx 1/2$ for the SM
Higgs of $m_h~=~100$~GeV and for $m_b~\approx~4.6$~GeV\@.
The neutralinos then subsequently decay to the hidden sector through
the couplings in Eq.~(\ref{eq:9}).  The decays are illustrated in Fig.~\ref{f.np}.

\begin{figure}[t]
 \hspace{-1cm}
\hbox{
\includegraphics[width=1.08\textwidth]{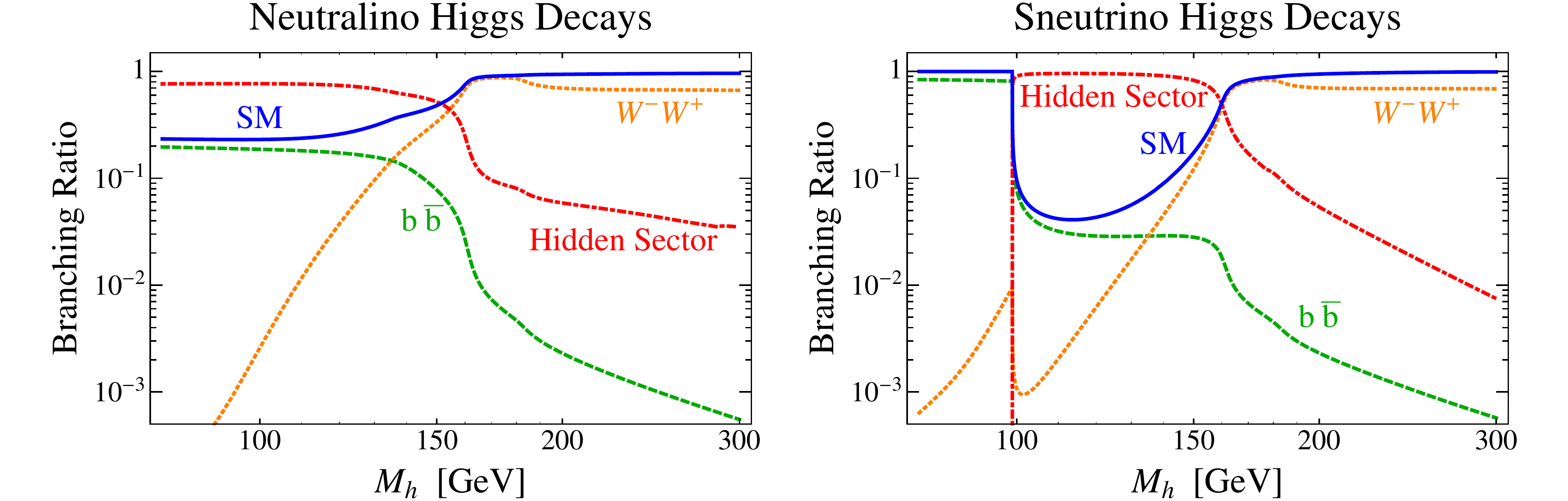}
}
\caption{\small \setlength{\baselineskip}
{.9\baselineskip} Higgs branching ratios for the neutralino and sneutrino channels.  Each plot shows the total Higgs branching ratios to the SM and hidden sector, as functions of the Higgs mass.  The SM width is dominated by the branching fractions to $b \bar b$ and $W^+ W^-$, which are also shown separately.   The parameters are fixed according to the benchmark models of Section \ref{subsec:benchmarks}.  For each model, the Higgs decays dominantly to the hidden sector below the $W^+ W^-$ threshold, and a 100 GeV Higgs satisfies the LEP constraint $\mathrm{BR}~(h\rightarrow~b \bar b)~<~0.2$.  The Higgs widths to SM states are taken from {\tt HDECAY}~\cite{Djouadi:1997yw}. }
\label{fig:HiggsBR}
\end{figure}

Even if $m_{N1} < m_h/2$ (which is fairly easy to arrange) it is not
automatic that the Higgs decays to neutralinos with a sizable
branching fraction.  Indeed, the coupling $g_{h11}$ vanishes if the lightest
neutralino is a pure gaugino or a pure Higgsino.  The decays into
neutralinos are therefore relevant only if the latter is a mixture of gauginos
and Higgsinos.  LEP constraints on a light SM Higgs require that
\begin{eqnarray}
  \label{eq:8}
  \Gamma (h\to b \bar b)/\Gamma(h\to \ti N_1 \ti N_1) \simlt 0.25
\end{eqnarray}
This is possible for example, if the lightest neutralino is dominantly
a bino, with a $10\%$ Higgsino fraction.   A corollary is that the
second visible neutralino and the lightest visible chargino cannot be
arbitrarily heavy as otherwise the mixing angles $c_{U,D}$ would be
suppressed.  Thus there is tension between Eq.~(\ref{eq:8}) and the
LEP  constraints on light charginos and the Tevatron constraints on trilepton signals
from decays of the second to lightest neutralino\footnote{Although the standard supersymmetry
  searches are not necessarily sensitive to these models, we require the charginos to be heavier than the LEP reach $\sim 100$~GeV.}.
Nevertheless, numerically one can find large portions of the
parameter space where the Higgs decay into the lightest neutralino
dominates, and at the same time the lightest chargino and the second
neutralino are significantly heavier than 100 GeV, and thus beyond LEP reach.   Trilepton constraints from the Tevatron, on the other hand, are model-dependent, because they depend on the branching fraction to trileptons and their kinematics.
In Section \ref{subsec:benchmarks}, we will consider an example spectrum that is not constrained by trilepton searches.

  Interestingly, all
the viable points we have found correspond to the lightest visible
neutralino mass below 40 GeV (this is consistent with the results of
\cite{linda}).  In consequence, the neutralino channel is constrained
by LEP-1 searches  (the $Z$ boson can decay to neutralinos via its
Higgsino component), but in the following we show that all existing
experimental constraints can be satisfied.  In the left plot of
Fig.~\ref{fig:HiggsBR} we show the Higgs decay branching fractions as
a function of the Higgs mass for the neutralino channel.  The
benchmark parameters of the specific model used are described in
Section~\ref{subsec:benchmarks}.

\subsection{Sneutrino Channel}
\label{sec:sneutrino-channel}

\begin{figure}[t]
\hbox{
\includegraphics[width=0.3\textwidth]{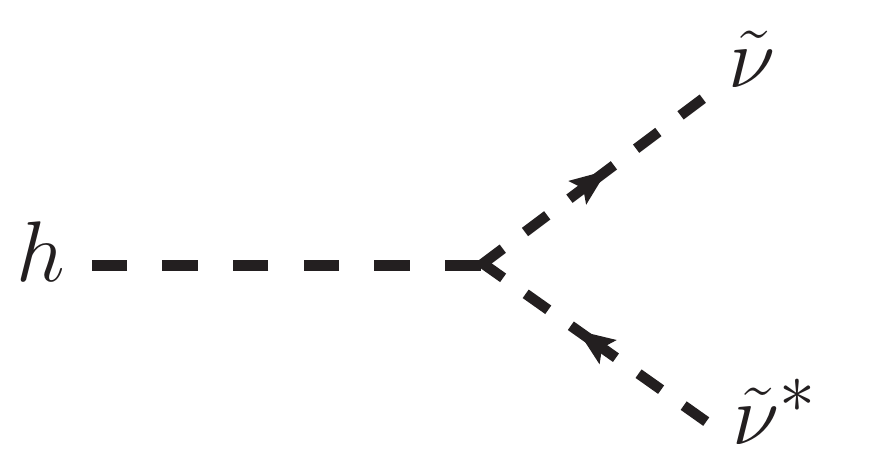}
\qquad 
\qquad 
\includegraphics[width=0.6\textwidth]{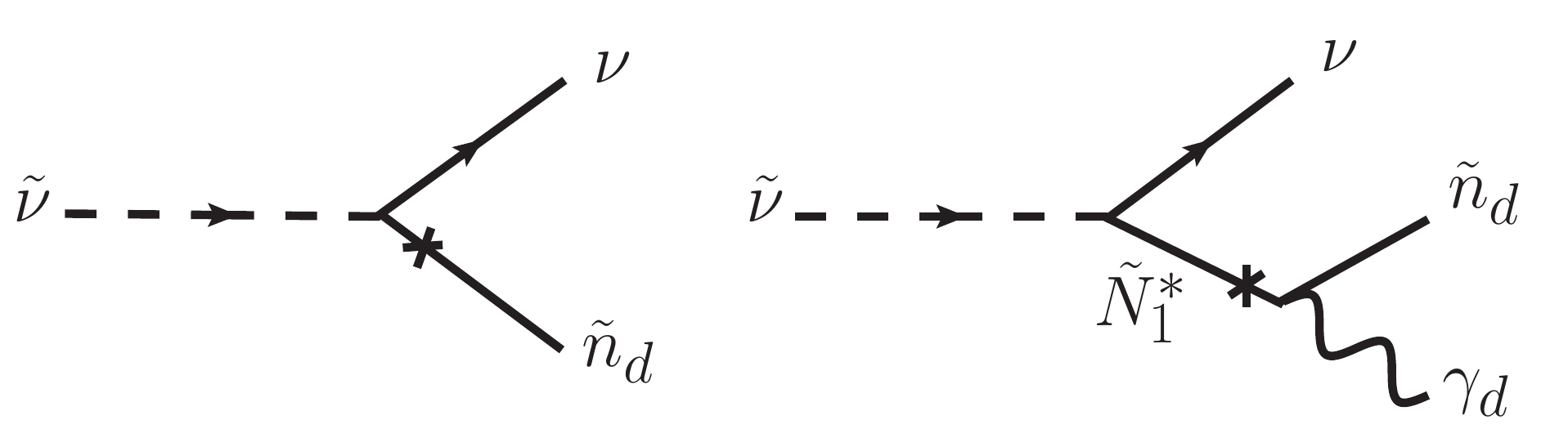}
}
\caption{\small \setlength{\baselineskip}
{.9\baselineskip} Diagrams relevant for Higgs cascade decays into the
  hidden sector via the sneutrino channel.  The {\it left} diagram shows
  the Higgs decaying into a sneutrino pair.  The {\it right} two diagrams are
  examples of two and three body decays of the sneutrino into hidden
  particles plus a SM neutrino.  The neutrino production implies an irreducible missing energy component for this channel, which is constrained by LEP $h \rightarrow E\!\!\!/_T$ searches, as discussed in Section \ref{subsec:ConstrainObservables}.}
\label{f.snup}
\end{figure}

The MSSM contains another class of neutral supersymmetric particles --
the sneutrinos -- which are the scalar partners of the SM left-handed
neutrinos.  If at least one of the sneutrinos is lighter than half the
Higgs mass, another channel is open for the Higgs to decay into the
hidden sector.  In the MSSM, the Higgs couples to the sneutrinos
through the $SU(2)_W\times U(1)_Y$ D-terms,
\begin{eqnarray}
  \label{eq:12}
  V_D \supset \frac{1}{8}\left(g^{\prime2}+g^2\right)\left(|H_u|^2 -
    |H_d|^2-|\tilde\nu_i|^2+...\right)^2\,,
 \end{eqnarray}
 where $g^\prime$($g$) is the hypercharge (weak) gauge coupling and the
 ellipses stand for additional terms which involve the sleptons and
 additional MSSM fields.  The above term induces a tri-linear coupling
 between the lightest MSSM Higgs and two sneutrinos,
\begin{eqnarray}
  \label{eq:11}
  {m_Z^2 \over v} \sin(\alpha+\beta)\,  h \, \ti \nu^\dagger\,  \ti \nu \,,
\end{eqnarray}
 where 
the resulting Higgs decay width is 
\begin{eqnarray}
  \label{eq:13}
 \Gamma(h\to \ti \nu \ti
\nu) = {m_Z^4\over 16 \pi m_h v^2} \sin^2(\alpha+\beta) \left (1 - 4
  { m_{\ti \nu}^2 \over m_h^2} \right )^{1/2}   \,.
\end{eqnarray}
The Higgs-sneutrino coupling is typically large, thus the $h\to \tilde \nu \tilde \nu$ decay
normally dominates over 
$h\to b\bar b$
as soon as it is kinematically allowed.  This is shown in the right plot
of Fig.~\ref{fig:HiggsBR} for the benchmark model described in
Section~\ref{subsec:benchmarks}.  
The sneutrino
cannot be lighter than $m_Z/2$ as this is excluded by the LEP-1
measurement of the $Z$-width.  This leaves the window $m_Z/2 < m_{\ti
  \nu} < m_h/2$ where the decay into sneutrinos can 
potentially lead to a hidden and light Higgs. 

Much as in the neutralino LVSP case, the sneutrino is not stable and
hence the Higgs decaying into sneutrinos is not invisible. 
There are two
ways for the sneutrino to decay into the hidden sector,  both illustrated in
Fig.~\ref{f.snup}.  Since the
bino is heavy in this scenario, Eq.~(\ref{eq:10}) induces a three-body
decay into a neutrino, a hidden neutralino and a hidden boson (scalar or  photon).  Conversely, the sneutrino can
decay directly into a SM neutrino and a hidden neutralino, through  the interaction in
Eq.~(\ref{eq:10}), however, the coupling is suppressed by an additional
$\mhid/m_Z$ factor.  As a consequence, the 3-body decay is dominant,
unless the lightest visible neutralino is significantly heavier than
100 GeV\@.  The sneutrino decay may be followed by a cascade in the
hidden sector leading to a final state with a number of visible
leptons and missing energy from the true LSP in the hidden
sector.  Due to the SM neutrino in the final state, the
sneutrino channel is characterized by more missing energy than the
neutralino channel.

As we will discuss below, the sneutrino channel, together with the minimal $U(1)_d$ hidden sector of Section \ref{sec:minimal-model}, suffers from considerable tension with several LEP searches. 
Two reasons are the typically larger missing energy
in the Higgs decays through the sneutrino channel and the independent 
sneutrino production cross-section through off-shell $Z$'s, which is comparable in rate to Higgs-strahlung.  An extended hidden sector, with additional cascades, can resolve this tension.  We return to these
issues in Section~\ref{sec:benchmark-models}.

 \subsection{Singlet Channel}
 \label{sec:singlet-channel}
 
Finally, new Higgs decay modes are possible, if there are additional mediators that
couple both to the Higgs and to the hidden sector.  A simple
example is constructed starting from the NMSSM in which an additional
singlet, $S$, couples to the Higgs doublets.  $S$ obtains a VEV,
thereby ameliorating the so called $\mu$-problem.  To enable the Higgs
decays, consider, as an example, the following superpotential,
\begin{eqnarray}
  \label{eq:14}
   W_{\rm singlet} = S\left (y \,\chi \,\bar \chi + \lambda \, H_u H_d \right)+ \kappa_1 \, \bar
   \chi \, h_1^2 + \kappa_2 \, \chi \, h_2^2 \,.
\end{eqnarray}
Here $\chi$, $\bar\chi$ are chiral superfields with charges $\pm2$
under $\Gd$ and $h_{1,2}$ are the two hidden Higgses.  Once the visible
Higgses and the singlet acquire VEVs, masses for $\chi$, $\bar\chi$ are
generated: the fermionic degrees of freedom get a mass $y \la S \ra$,
while the scalar masses are split by the $F$-term of $S$, $m_{\chi}^2
= y^2 \la S\ra^2 \pm y \, F_S$.  If the lightest scalar state is lighter
than $m_h/2$, the Higgs can decay into a pair of these fields with a
large branching fraction, as long as the couplings $y$ and $\lambda$
are sizeable.  Quite generically, the branching fraction into the
hidden sector is close to unity.  The branching fraction for the
benchmark model described in Section~\ref{sec:benchmark-models} is
shown in the right plot of Fig.~\ref{fig:SingletBranch}.  

\begin{figure}[t]
\hbox{
\includegraphics[width=0.3\textwidth]{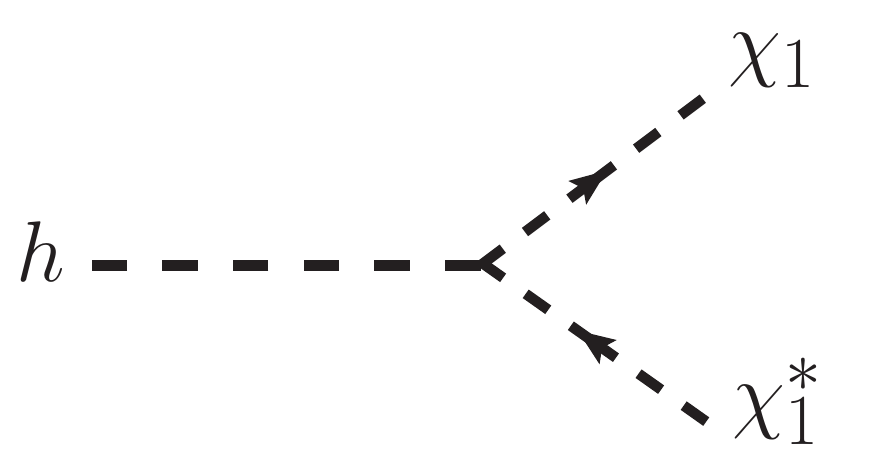}
\qquad 
\qquad 
\includegraphics[width=0.6\textwidth]{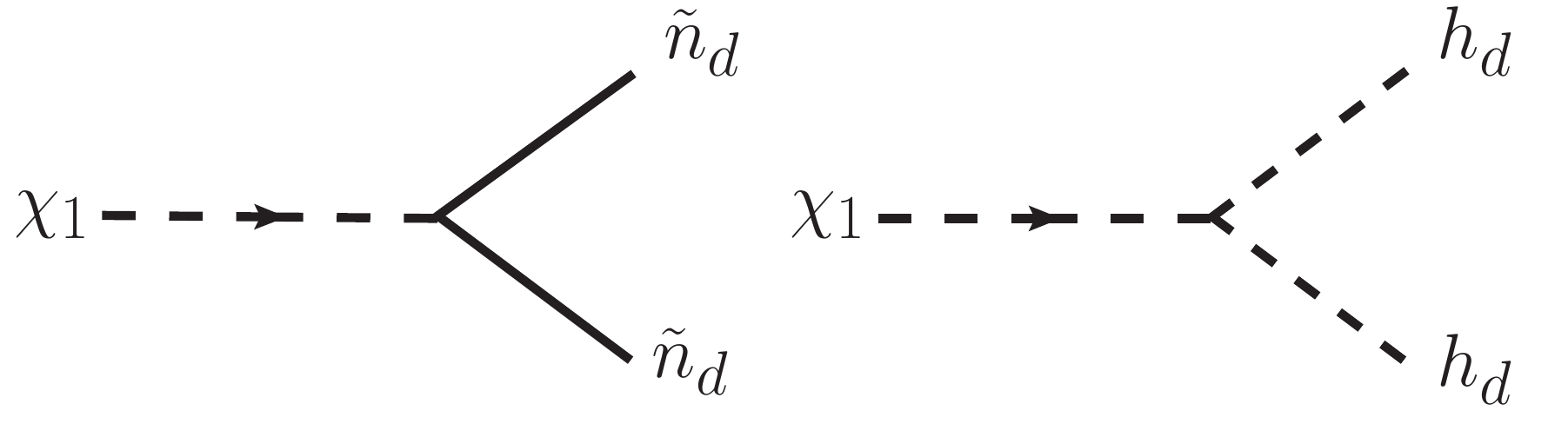}
}
\caption{\small  \setlength{\baselineskip}
{.9\baselineskip} Diagrams relevant for Higgs cascade decays into the
  hidden sector via the singlet channel of Eq.~\eqref{eq:14}.  The scalars that couple to the Higgs are split by the NMSSM singlet $F$-term, such that the lighter one, $\chi_1$, is easily lighter than half the Higgs mass.  The scalars couple to light hidden sector Higgses through the superpotential, and the {\it right} two diagrams are examples of the resulting decays of $\chi_1$ into hidden fermions and scalars.}
\label{f.sinp} 
\end{figure}

 Subsequently, the $\chi$ states
decay via the $\kappa_{1,2}$ couplings into hidden Higgsinos and Higgses,
which finally decay to SM fermions.  Once again the final state of
the Higgs decay is a number of leptons plus missing energy.  The
virtue of this model is that it requires a minimal deformation of the
NMSSM (or other variants which address the $\mu$-problem), and is by
and large independent of the visible spectrum.  It can therefore
accommodate heavy SM superpartners which are beyond the LEP and Tevatron reach,
while allowing the Higgs to decay dominantly into the hidden sector.  Such
a model is therefore in principle less constrained by existing
searches.

\begin{figure}[t]
\bc
\includegraphics[width=0.55\textwidth]{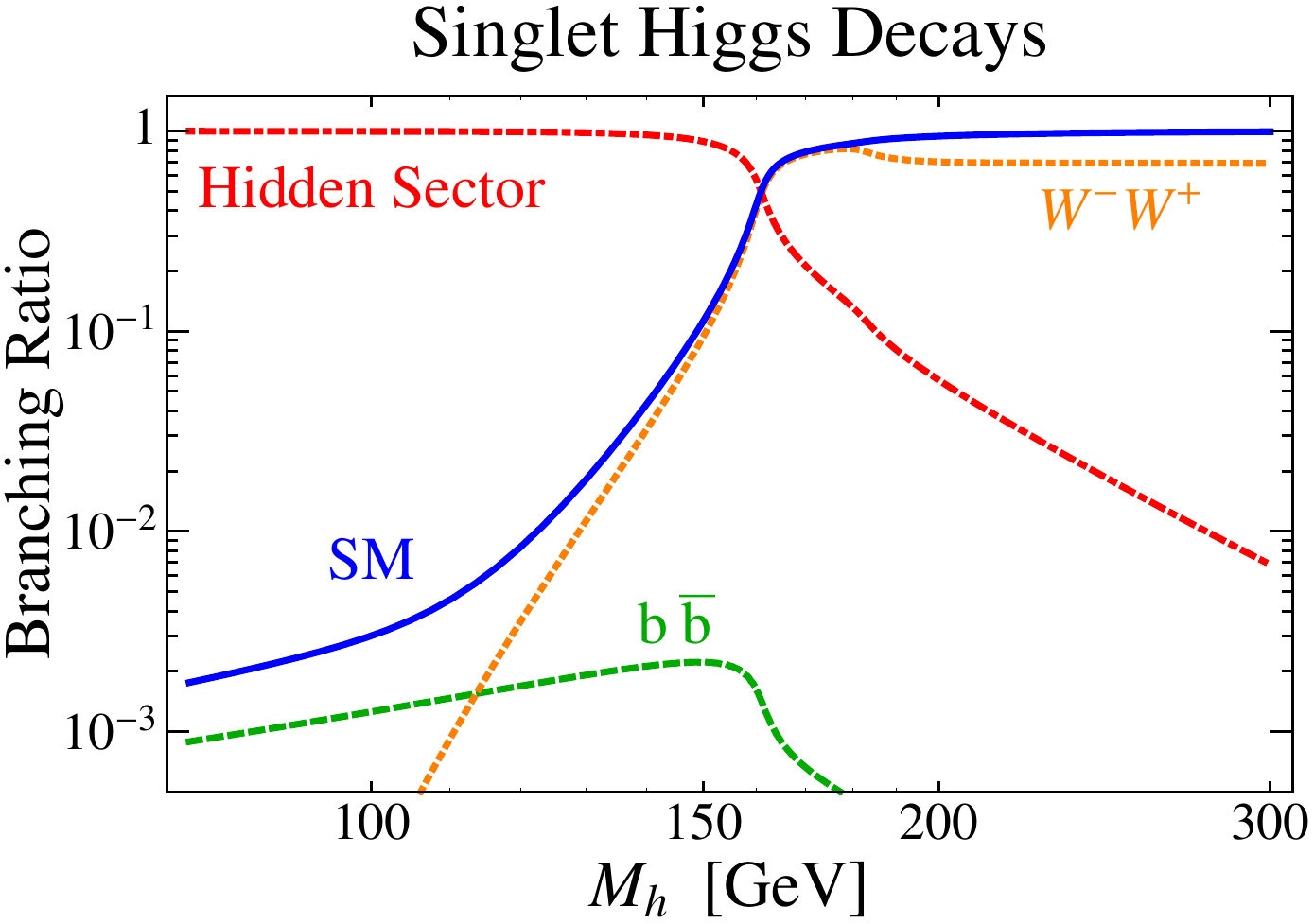}
\ec
\caption{\small  \setlength{\baselineskip}
{.9\baselineskip}Same as in Fig.~\ref{fig:HiggsBR}, for the singlet channel, with the parameters fixed according to the benchmark of Section \ref{subsec:benchmarks}.}
\label{fig:SingletBranch}
\end{figure}

\section{Collider Phenomenology}
\label{s.cp}

In order to establish models where the Higgs and possibly SUSY are hidden, we must understand how the phenomenology of this scenario is experimentally constrained.  In this section, we describe the collider physics of a Higgs decaying to lepton jets.  We begin, in Section \ref{sec:lept-neutr-jets}, with a general description of lepton jets and neutralino jets, which are a particular subclass defined below.  Then, in Section \ref{sec:exper-observ}, we describe the most prominent experimental variables of this scenario, defining a space of observables.  After discussing the relevant LEP and Tevatron searches in Section~\ref{s.ec}, we continue in Section~\ref{sec:benchmark-models} by identifying the allowed region in this space of experimental observables.  We then construct benchmark models that reside in this allowed region. 

 \subsection{Lepton Jets and Neutralino Jets}
 \label{sec:lept-neutr-jets}
 The Higgs can avoid detection at LEP and the Tevatron if it decays into final states that are relatively unconstrained by existing searches.  Here we focus on one such class of final state objects: {\it lepton jets}, which are defined as high-multiplicity clusters of boosted and collimated leptons \cite{nn, itay, baumgart}.  Lepton jets can be produced when the Higgs decays through low lying hidden sector states with masses (in
 particular the hidden gauge boson) below $\lesssim 1$ GeV.  These hidden sector states then decay to leptons.  The
 picture is the following.  First, the Higgs boson is produced, alone or
 in association with $Z$ or $W$, and decays into a pair of SM superpartners or singlets.  We discussed three such channels in \sref{p}. 
 The pair promptly decays into the hidden sector, cascading through hidden sector interactions to produce
 highly boosted hidden scalars, photons and neutralinos.  These appear in the detector either as missing energy or produce
 boosted leptons which populate lepton jets.  An
 example of such a Higgs cascade decay is depicted in \fref{shd}.  

 The
 number of lepton jets per Higgs decay depends on the cascading spectrum, and the resulting topology is easily deduced by recalling that particles produced at rest decay to well-separated objects while boosted particles decay to collimated objects.  For example, if the Higgs decays to two weak scale SM superpartners or singlets, each will decay to well-separated hidden particles that seed distinct lepton jets.  Subsequent decays will be collimated, and the event topology can contain as many as four lepton jets.  Alternatively, for example, if the Higgs decays to two very light neutralinos, $m_{\tilde N_1} \lesssim 10$~GeV, then each neutralino's decay products will be clustered, and the event topology will contain two lepton jets, one for each neutralino.  
 In this situation, we refer to the lepton jets as {\it neutralino jets}.  We consider an example with neutralino jets in Section~\ref{subsec:benchmarks}.

\subsection{Experimental Observables} 
 \label{sec:exper-observ}

Concentrating on the 
 lepton jets, we
identify the following  relevant collider variables,

\begin{itemize}
\item {\bf Visible Final States: Electrons vs. Muons} 
\begin{figure}[t]
\bc
\includegraphics[width=0.6\textwidth]{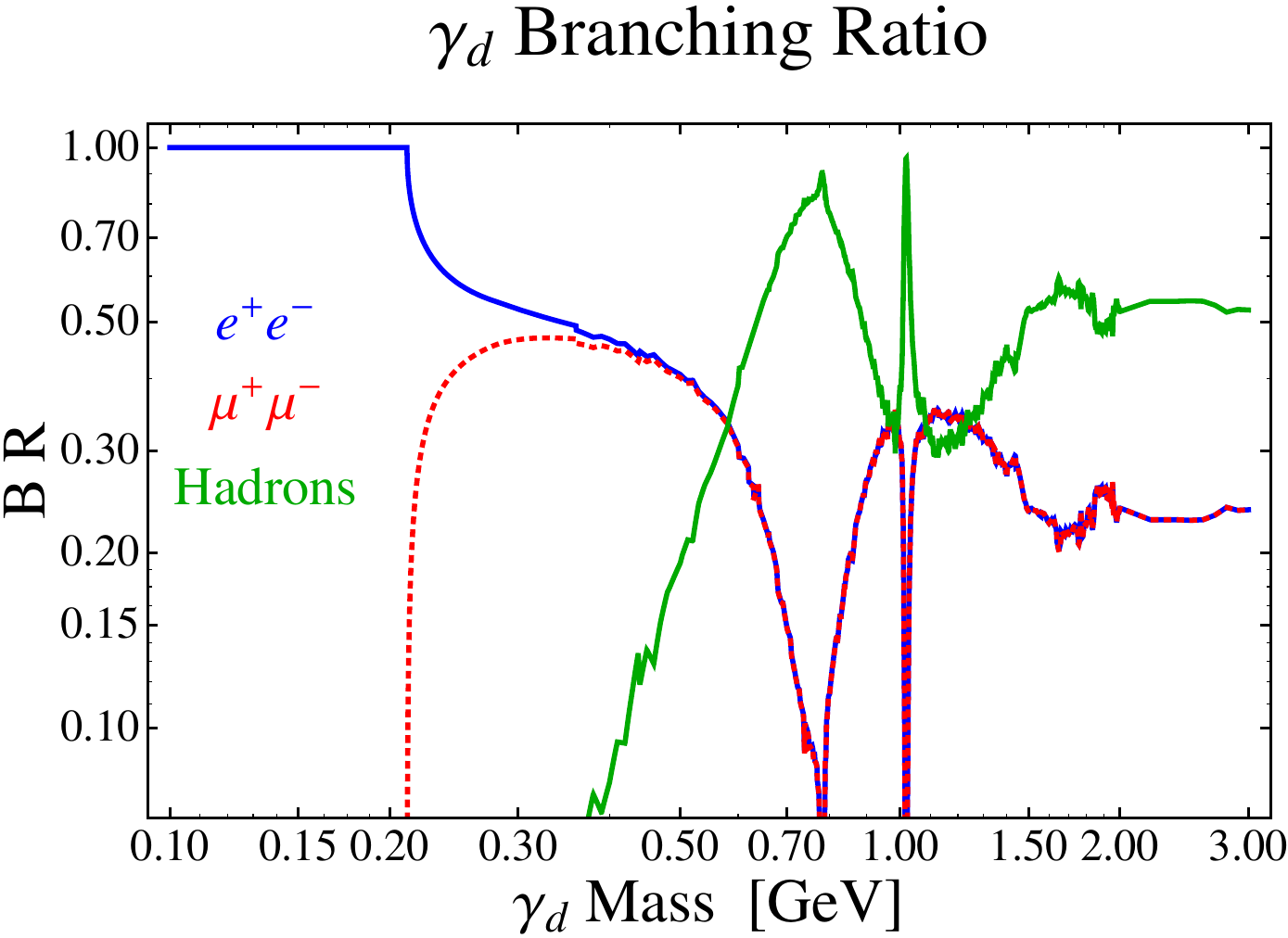}
\ec
\caption{\small  \setlength{\baselineskip}
{.9\baselineskip}
Hidden photon branching ratios to electrons, muons, and hadrons through the electromagnetic current, as a function of the hidden photon mass.  The hadronic branching ratio is derived from the measured $R \equiv {\rm BR}(e^+e^- \to {\rm had})/{\rm BR}(e^+e^- \rightarrow
{\rm \mu^+ \mu^-})$ \cite{Amsler:2008zzb}.  We see that for $\mhid~\lesssim~500$~MeV, the hidden photon decays dominantly to leptons, including muons for $\mhid >211$~MeV.}
\label{f.nl}
\end{figure}
\\
The hidden photon decays through the kinetic mixing, Eq.~\eqref{e.km}, into all kinematically allowed SM states with electric charge.  Thus, the mass of the hidden photon is the only parameter
controlling which visible particles appear at the end of the Higgs
cascade decay.  For $\mhid < 2 \,m_{e}$ the
hidden photon is stable on collider scales,  which would amount to a purely
invisible Higgs signature.  This scenario is strongly constrained, as we review in Section \ref{s.ec}, and we do not consider this possibility here.
For $2 \,
m_e < \mhid < 2 \,m_{\mu}$ the hidden photon decays exclusively into
electrons.  For $2 \, m_{\mu} < \mhid < 2 \, m_{\pi^\pm}$ the hidden photon
decays into a pair of muons or electrons, roughly in equal amounts.
For $\mhid > 2 \, m_{\pi^\pm}$ the branching fractions are determined by
$R \equiv {\rm BR}\, (e^+e^- \to {\rm had})/{\rm BR}\, (e^+e^- \rightarrow
{\rm \mu^+ \mu^-})$~\cite{Meade:2009rb, Batell:2009yf} which is
measured experimentally~\cite{Amsler:2008zzb}.  As long as $\mhid \lesssim 400$ MeV, BR ($\zd
\to \pi^+ \pi^-$) is less than $10\%$ and can be safely neglected, as
we do below.  The branching fractions are shown in Fig~\ref{f.nl}.
 
\item {\bf Lepton Multiplicity} \\
  This observable is extremely sensitive to the details of the hidden
  sector spectrum.  One important factor is the identity of the
  lightest hidden neutralino.  Since the visible bino couples to
  hidden Higgsinos, see Eq.~(\ref{eq:9}), model realizations where the
  hidden bino is lighter than the hidden Higgsinos have longer cascades, and
  therefore tend to produce more visible leptons.  Another crucial
  factor is the ratio of the masses of the lightest hidden scalar and
  hidden photon.  When $\mhhid < \mhid$, the hidden Higgs, $\hd$, dominantly decays to 2 leptons at one-loop \cite{Batell:2009yf},
  and is
  stable on collider scales.   On the other hand, for $\mhhid > \mhid$
  the 3-body decays with one on-shell hidden photon are allowed, which
  leads to prompt decays of $\hd$ into 4 leptons, as long as the mixing
  parameter $\eps$ is not too small.  The spectrum of the other hidden scalars is also important.  For example, when $m_{\Hd} > 2 \, \mhhid$ the dominant
  decay mode of $\Hd$ is $\Hd \to 2 \, \hd \to 8 \, l$, while for $\mhid <
  m_{\Hd} < 2 \, \mhhid$ the 4-lepton decay via the hidden photons
  dominates.  
  Depending on the mass spectra, the average lepton multiplicities can thus range from zero to a few tens of leptons
  per Higgs decay.  Going beyond the minimal $U(1)_d$ model, for example
  by making the hidden group non-Abelian, only increases the number of
  possibilities.  Additionally, if the hidden gauge coupling is
  sizeable ($g_d^2/4 \pi \gtrsim 0.1$), hidden sector
  showering also increases the number of
  leptons~\cite{Meade:2009rb,itay}.  Example lepton
  multiplicity distributions are given in \fref{met}.
\begin{figure}[t]
\hbox{
 \hspace{-1cm}
\includegraphics[width=1.08\textwidth]{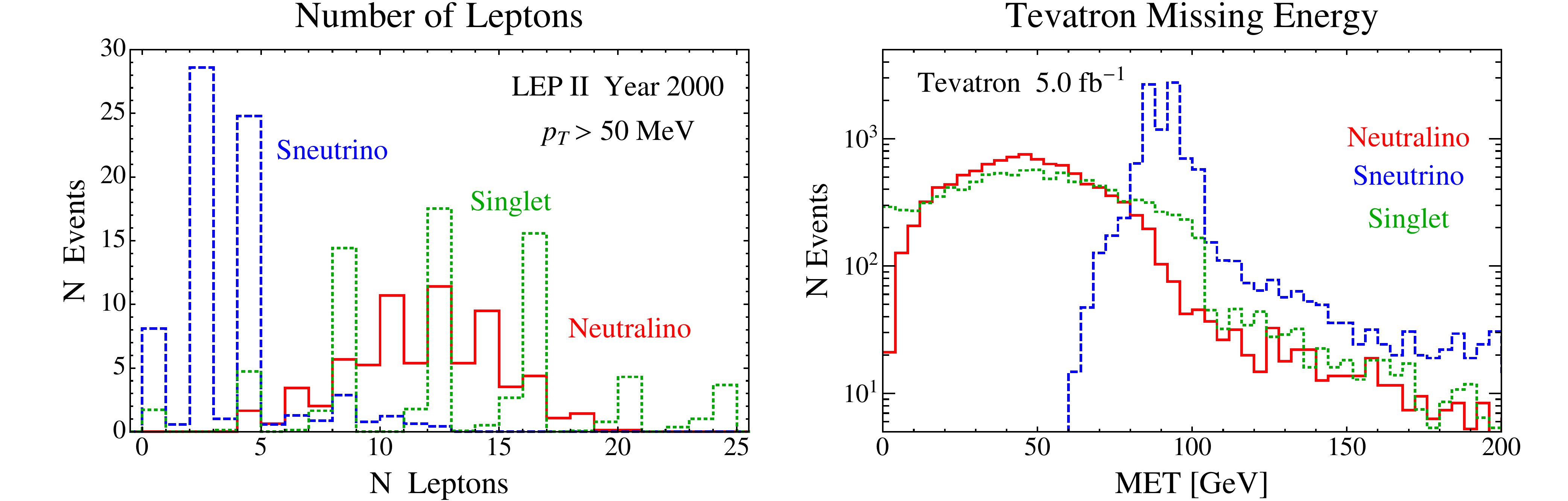}
}
\caption{\small  \setlength{\baselineskip}
{.9\baselineskip}
The lepton multiplicity and missing transverse energy distributions of Higgs decays for the three benchmarks of section \ref{subsec:benchmarks}.  The {\it left} panel shows the multiplicity of leptons with $p_T > 50$ MeV, which is roughly the threshold for detecting tracks at LEP\@.  The event counts correspond to the data collected in the year 2000 at LEP-2, corresponding to $\mathcal{L} \approx 214\unit{pb}^{-1}$ at $\sqrt s~=~205-207$~GeV\@.  The {\it right} panel shows the missing energy distribution at Tevatron Run II, $\sqrt s = 1.96$ TeV, and the event counts correspond to $\mathcal{L} = 5 \unit{fb}^{-1}$.  The sneutrino benchmark has the most missing energy because of the irreducible missing energy carried by the neutrinos.
}
\label{f.met}
\end{figure}
\item {\bf Missing Energy} \\
  The average missing energy per Higgs decay is also very sensitive to the
  hidden spectrum.
  The missing energy can range anywhere from being very
  small, less than 10 GeV, to where it dominates over visible energy.
  The most important factor determining the amount of missing energy
  is how many hidden particles
   are
  collider-stable.  Furthermore, missing energy depends on
  the Higgs decay channel into the hidden sector.  For the sneutrino
  channel the amount of missing energy is typically larger, because a
  hard neutrino is emitted when the sneutrino decays into the hidden
  sector.  For typical hidden spectra the average missing transverse
  energy per Higgs decay is on the order of 10-50~GeV, and displays a
  large variation on an event-by-event basis, see \fref{met}.
 
\item {\bf Event Topology: Number of Lepton Jets} 
 \begin{figure}[t]
\bc
\includegraphics[width=.95\textwidth]{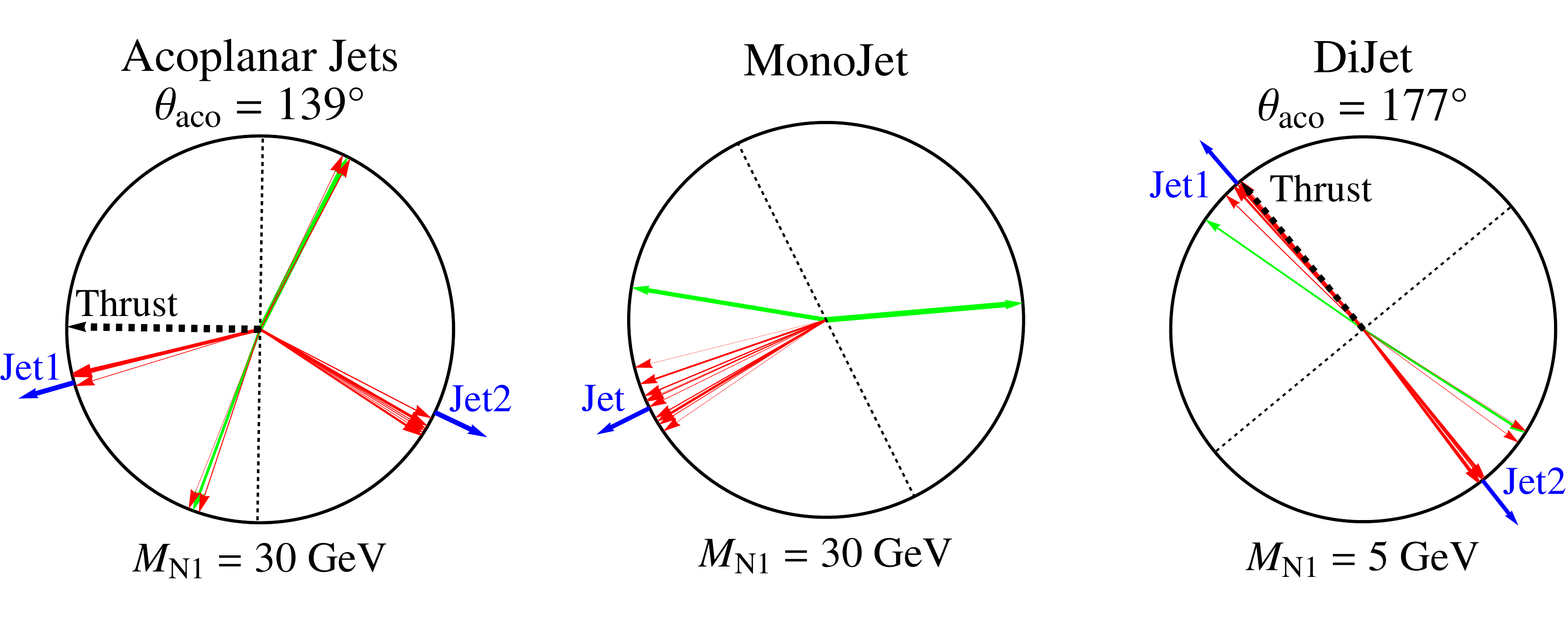}
\vspace{-0.75 cm}
\ec
\caption{\small  \setlength{\baselineskip}
{.9\baselineskip}
Transverse event displays for $Z$ decays to neutralinos at LEP-1.  The red vectors show the transverse momenta of leptons and the green vectors correspond to long-lived hidden sector particles that escape the detector as missing energy.  The {\it left} display shows a 4-lepton-jet event, with $m_{\tilde N_1} = 30$ GeV, that would have been detected by the ALEPH acoplanar jet search~\cite{Aleph_aco}.  The thrust (black) is used to define two jets (blue).  In an acoplanar event, the jets are separated by $\theta_{\rm aco} < 175^\circ$ in the transverse plane.  The {\it central} display shows an event that would have been detected by the ALEPH monojet search~\cite{Aleph_mono} because all visible energy falls in the same hemisphere.  Here, both neutralinos produce exactly one visible lepton jet in the same hemisphere, balanced by missing energy.  Both acoplanar jets and monojets are avoided if $m_{\tilde N_1} \lesssim 5$ GeV with both neutralinos decaying (partly) visibly,  as in the {\it right} display.  The topology of the back-to-back neutralino jets mimic the hadronic dijet background.
}
\label{f.et}
\end{figure}  
\\
  The directions of final state lepton momenta are not distributed
  randomly.  
  Hidden sector particles are produced with large boosts, and their clustered decay products populate distinct lepton jets containing highly collimated leptons.  The final state topology is characterized by the number of lepton jets, which depends on the spectrum and first steps of the cascade decay.    A two-jet topology
  arises when the Higgs decays directly into two GeV scale objects, or
  if the superpartners decay into one hidden particle that decays visibly along with other invisible particles.  The former possibility is most easily realized
  with the neutralino jets discussed in Section \ref{sec:lept-neutr-jets}.  Indeed, if the lightest
  MSSM neutralino mass is below $10$~GeV (which is possible without
  compromising naturalness, and without conflicting experiment), its
  visible decay products form a jet along its direction of motion.
  On the other hand, when the Higgs decays into neutralinos heavier
  than 10 GeV, or when the three-body decays of the sneutrino channel
  dominate, the event topology will contain 3 or more jets.  Finally,
  if one of the two primary Higgs decay products (neutralinos, sneutrinos or
  hidden fields) decays invisibly, while the other decays into one lepton
  jet, the final state will display a monojet topology.  All three
  possibilities are shown in Figure~\ref{f.et}.

\item {\bf Lepton Isolation} 
   \begin{figure}[t]
\bc
\includegraphics[width=0.6\textwidth]{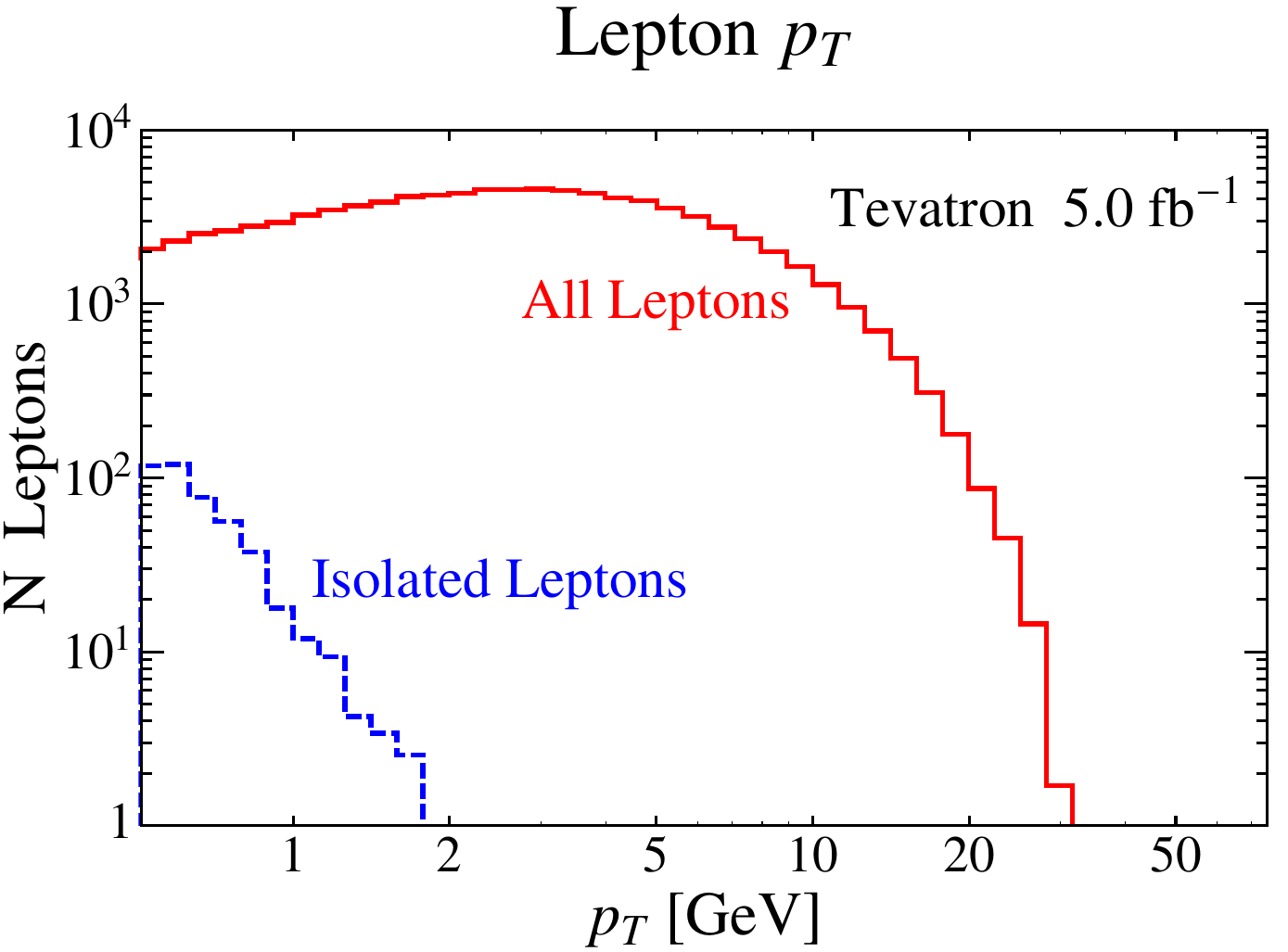}
\ec

\caption{\small  \setlength{\baselineskip}
{.9\baselineskip}
The $p_T$ distributions of all leptons (solid, red) and isolated leptons (dashed, blue) produced in gluon fusion Higgs decays at Tevatron Run II, $\mathcal{L} = 5\unit{fb}^{-1}$, for the singlet benchmark of section \ref{subsec:benchmarks}.  Tevatron searches for trileptons and like-sign dileptons impose strict isolation requirements in order to fight the SM heavy flavor and in-flight-decay backgrounds.  For this plot, we use a track-based isolation definition where the scalar $p_T$ sum of all other tracks in a cone surrounding the lepton, $\Delta R < 0.4$, must not exceed $10\%$ the $p_T$ of the lepton \cite{Abulencia:2007rd}.  We see that Higgs decays to lepton jets produce few isolated leptons, all of which are soft.}
\label{f.LepIso}
\end{figure}
 \\
  Yet another consequence of their boosted origin is that lepton jets are narrow, with constituent leptons that are not isolated.  Tevatron searches for new physics that produces leptons, such as the trilepton and like-sign dilepton searches discussed in Section~\ref{sec:tevatron-searches}, require isolated leptons in order to fight the backgrounds from heavy-flavor decays and in-flight meson decays.  These searches typically require leptons to be isolated within cones of $\Delta R < 0.4$.  Some searches require total isolation in this cone \cite{Cdf_3luni}, while other searches put limits on the energy deposition in the calorimeters \cite{Cdf_3llow}, or the sum of track $p_T$ \cite{Abulencia:2007rd}.  Lepton jets violate these isolation definitions because each hidden
  photon decays into a pair of close leptons separated by $\Delta R
  \lesssim 0.1$.  Thus, almost all leptons in the final state have
  at least one companion track within the isolation cone\footnote{Isolated leptons are sometimes produced when soft leptons are emitted at wide angles from the rest of the lepton jet or when nearby leptons are too soft to reach the calorimeter (e.g. $p_T
    \leq 0.5$ GeV at Tevatron), and are therefore interpreted as missing energy.}.  Furthermore, the
  bulk of our parameter space leads to large lepton multiplicities
  with numerous leptons in a $\Delta R \sim 0.1$ cone, further
  spoiling the isolation.  In \fref{LepIso} we show the distribution
  of the total number of leptons as a function of the transverse
  momenta together with leptons that satisfy the track-based isolation of Ref.~\cite{Abulencia:2007rd}.  One can see that only a small fraction
  of the leptons are isolated, and these
  are typically soft leptons that would not pass the $p_T$ cuts for
  most new physics analyses.  The other isolation definitions give similar results.

\item {\bf Displaced Vertices} \\
  A final feature of this class of models is the possible existence
  of displaced vertices.  For instance, displaced vertices are produced if
  the kinetic mixing, $\eps$,  is small enough. The hidden photon decay
  length scales as $1/\eps^2$, so that for hidden photon mass $\mhid \simeq
  0.1$~GeV, displaced vertices show up for $\eps \lesssim 10^{-5}$.  
  Displaced vertices can also appear for a subset of final state
  particles, in three-body decays, if 
  mass splittings in the
  hidden sector are small compared to the GeV scale.  This can occur
 without tuning. For
  example, the splitting between two hidden neutralinos can be
  naturally small if they are both Higgsino-like.  The presence
  of displaced vertices would avoid most of the existing LEP and
  Tevatron constraints, as such signal events would not be selected by
  most analyses.  Higgs decays with displaced vertices were studied
  in~\cite{Strassler:2006ri, Graesser:2007yj}.  To keep our discussion simple we do
  not consider them in this paper.  Nonetheless, one
  should keep in mind that this possibility could be realized in nature and would
  relax the constraints on new physics.  Conversely, designing
  collider search strategies that would be sensitive to a Higgs decaying
  to lepton jets with displaced vertices is an important task that we
  postpone for future work.

 \end{itemize}

 \section{Can Lepton Jets Really Hide the Higgs?}
\label{sec:crazy}

Before considering specific searches, it is natural to ask whether Higgs decays to lepton jets would have been trivially seen at LEP and/or the Tevatron. 
In
 other words, one might think that it is naive to imagine that such
 spectacular events can remain undetected.  To address this worry let us first consider LEP-2, which sets the strongest limits on a light Higgs.
 The production cross-section, for a $100$ GeV Higgs, is around $0.3 -
 0.4$ pb.  The total luminosity at LEP-2 is on the
 order of $450$ pb$^{-1}$, per experiment.  The number of Higgs-strahlung events before
 cuts is therefore~$\sim 130$.  Preliminary cuts typically reduce the number of events by an order of magnitude, leaving only a handful of events and 
making detection nontrivial without a dedicated search. 
 Furthermore, due to the relatively poor quality of the hadronic
 calorimeters at the LEP experiments, hadronic activity is typically
 identified by the number of tracks.  For this reason, even at LEP-1
 where the neutralino channel allows for as many as $10^4$ lepton jet
 events, these events are easily hidden beneath the large hadronic
 background.
 
 At the Tevatron, Higgs production is dominated by gluon fusion with a
 cross-section of $\sim 2$ pb for a $100$ GeV Higgs.  With $5$
 fb$^{-1}$ this implies $10^4$ Higgs events.  Of course, the large
 QCD background does not allow one to search for such events `by eye'.
 Even so, one may worry that the existence of many leptons
 would, naively, allow for an easy discovery.  The key point is that Tevatron searches for leptons require them to be isolated.
 This is true, in particular, for the Vista/Sleuth global searches which
 take a model-independent approach~\cite{VistaSleuth}.  The leptons produced by Higgs decays
 to lepton jets are typically not isolated, and thereby evade the standard
 searches.  The small number of Higgs events that do produce isolated leptons are buried beneath the QCD and electroweak backgrounds.

 \section{Experimental Constraints}
\label{s.ec}

The strongest constraints on the Higgs mass and decay branching
fractions come from LEP-2 searches.  These are obtained assuming the
Higgs is dominantly produced through the Higgs-strahlung process, $e^+
e^- \to Z h$, with the SM cross section.  Three well known results should
be kept in mind:
\begin{enumerate}
\item The most robust constraint on the Higgs mass comes from the model independent
  search by OPAL~\cite{Opal_general}, $m_h \geq 82$ GeV.  This
  bound is independent of the Higgs decay modes.
\item The mass of the SM Higgs is constrained to be $m_h\geq 114.4$
  GeV~\cite{Lep_smHiggs}.  This bound is obtained by studying the
  dominant $h\to b\bar b$ SM decay.  Conversely, the study can be
  interpreted as a bound on the $h\to b\bar b$ branching ratio, which
  for a $100$ GeV Higgs is bounded to be BR$(h\to b\bar b) \lesssim
  20\%$~\cite{Lep_4b}.
\item Finally, the invisible Higgs (where the Higgs decays exclusively into
  missing energy) must be heavier than 115 GeV~\cite{Lep_invisible}.
  Again, the bound can be interpreted also as the bound on the branching ratio of
  Higgs to invisible states.  For a $100$ GeV Higgs the bound is ${\rm
    BR}(h\to E\hspace{-.25cm}\slash) \lesssim 15\%$.
\end{enumerate}

The LEP collaborations have performed several searches for Higgs
decaying into 2 SM particles other than the $b$ quarks, always placing a
bound on the Higgs mass almost as stringent as the SM one,
see~\cite{Chang:2008cw} for a review. 
Higgs decaying into a larger
number of SM states is considerably less constrained, with the
exception of $h \to 4b$ and $h \to 4 \tau$ channels.  Thus, Higgs
decaying into high multiplicity final states offers a way to
circumvent the LEP limits, as long as 2-body decay channels are
sufficiently suppressed.
 
Higgs decaying to lepton jets has not been searched for at LEP or the
Tevatron.  In this scenario the Higgs mass  is in principle
constrained only by the OPAL model-independent limit.  
But the final states predicted by this scenario could well be picked up by some existing Higgs or new physics
searches.  This section provides an overview of the most relevant
searches at LEP and the Tevatron together with a brief explanation why they could be
sensitive to the Higgs decaying to lepton jets.

\subsection{LEP-1 Searches}
\label{sec:lep-1-searches}

The LEP-1 searches are relevant for the neutralino channel, Section \ref{sec:neutralino-channel}, because in
this scenario the lightest MSSM neutralino is necessarily lighter than
$m_Z/2$ \cite{linda}.  The $Z$ boson can then decay into a pair of
neutralinos, each of which decays via the hidden sector cascade into a
hidden neutralino and lepton jets.   Even though $BR(Z\to \ti N_1 \ti N_1)$
can be as small as $10^{-3}-10^{-4}$, this still leaves $10^{3}-10^{4}$
neutralino/lepton jets at LEP-1.  For the sneutrino channel, Section \ref{sec:sneutrino-channel}, the
measurement of the $Z$ width at LEP-1 constrains the 
branching ratio of $Z$
into any new particle to be smaller than $10^{-3}$ \cite{Lep_zpole}.
This immediately implies $m_{\ti \nu} > m_Z/2$.  For the singlet
channel, Section \ref{sec:singlet-channel}, the LEP-1 searches are not important since the hidden fields do not couple to $Z$ at leading order.
     
We have identified the following searches that could be sensitive to $Z\to \ti N_1 \ti N_1\to {\rm lepton ~jets}+\mE$ decays in our scenario

\begin{itemize}
\item  {\bf Monojets} \\
  In \cite{Aleph_mono} the ALEPH collaboration analyzed the so-called
  monojet events where no energy is detected in the hemisphere
  opposite to the direction of the total visible momentum.  This can
  happen, if the neutralinos are produced at rest or in those corners
  of the parameter space where the hidden cascade yields a low
  multiplicity of visible states, in particular, if one of the
  neutralinos 
   decays invisibly.  Conversely, these
  constraints are typically avoided, if the neutralino 
  decays to a large number of visible states.

\item  {\bf  Acoplanar Jets} \\
  Events can be clustered into two jets by summing the total momenta
  in each of the two hemispheres defined by the plane perpendicular to
  the thrust axis.  Acoplanar jets are then defined by requiring that
  the angle between transverse momenta of the two jets is smaller than, e.g.,
  $175^\circ$ \cite{Aleph_aco}.  In Ref.~\cite{Aleph_aco} the ALEPH collaboration
  searched for acoplanar jets accompanied by missing energy.  The
  signal events often contain a large number of charged tracks
  (especially if the model satisfies the monojet constraints).
  Therefore they may be efficiently picked up by this analysis, even
  though it was designed to search for hadronic events.

\item  {\bf Energetic Lepton Pairs}   \\
  Also in Ref.~\cite{Aleph_aco}, ALEPH made a search for energetic
  lepton pairs in hadronic events.  In the neutralino channel, Section \ref{sec:neutralino-channel}, two of
  the multiple leptons from the neutralino jet can readily meet the
  definition of the energetic pair.

\end{itemize}

\subsection{LEP-2 Searches}
\label{sec:lep-2-searches}

The following LEP-2 Higgs searches are sensitive to the events $e^+e^-\to Zh$ with $h\to {\rm ~lepton~jets}+\mE$ and could potentially constrain our scenario.  
\bi 
\item {\bf Flavor-independent Higgs} \\
  LEP has constrained the Higgs boson decaying into generic jets
  without relying on $b$-tagging.  These searches can be relevant since
  our signal often displays a two-jet topology.  The analysis of the
  OPAL collaboration \cite{Abbiendi:2003gd} is the most straightforward to
  interpret in our framework, because it does not rely on neural
  network techniques. 
 However the presence of missing energy makes the $H \to 2j$ searches less sensitive to the signal, as compared for example with the squark searches described later in this subsection.     
\item {\bf Invisible Higgs} \\
  Searches for the $Z h$ final state where $h$ decays entirely into
  missing energy have been performed by all collaborations
  \cite{Lep_invisible}.  
  The most relevant for us is the OPAL invisible Higgs search
  \cite{Opal_invisible}, because its visible mass cut is the least
  aggressive, $50{\rm ~GeV}< M_{\rm vis}< 120{\rm~GeV}$.  This search can
  strongly constrain our models, especially the sneutrino channel where the
  invisible energy fraction is typically larger due to the neutrinos
  produced by sneutrino decays.  This search can also pick up the signal events in the case of the neutralino and the singlet
  channel models, 
  if the associated $Z$ boson decays invisibly.
\item {\bf Higgs to $WW^*$} \\
  In Ref. \cite{Aleph_wwstar} the ALEPH collaboration 
  performed a
  search for $h\to W W^*$ decays
  in the context of
  fermiophobic Higgs models.  Leptonic decays of $W$ lead to final
  states with electrons/muons and missing energy, so that the ALEPH search
  targets final states similar to our signal --- lepton jets and missing energy.  Furthermore,
  the data sample is systematically  divided into distinct classes
  corresponding to different decay topologies of the $WW^*$ system.
 The $h \to WW^*$ search thus turns out to be very
  sensitive to Higgs decaying to lepton jets.
\item  {\bf Higgs to $4\tau$} \\
  Very recently, an analysis of the $h \to AA \to 4 \tau$
  decay was presented by the ALEPH collaboration \cite{kyle}.
  This search targets the case where the intermediate pseudoscalar $A$
  is very light, 10 GeV or less, in which case Higgs decays into two
  pairs of nearly overlapping $\tau^\pm$.  Since tau leptons decay
  into 1 or 3 charged particles most of the time, the analysis focuses
  on 2-jet events that contain 2 or 4 tracks per jet, while the
  associated $Z$ is assumed to decay invisibly or leptonically.  From
  our perspective, the analysis is very relevant, since 
  no $\tau$ identification is attempted, other than constraining
  the number of tracks.  Therefore,  Higgs-to-lepton jets signal might be picked
  up, 
  if the Higgs decays to a small enough number of
  charged states.
\end{itemize}
Apart from the Higgs searches, certain SUSY
searches could also be sensitive to the Higgs-to-lepton jet final state,
\begin{itemize}
\item {\bf R-parity Violation} \\
 If there is R-parity violating operator $LLE^c$ in the
  superpotential, the lightest neutralino or sneutrino decay to
  leptons and neutrinos.  In that context, ALEPH \cite{Aleph_rpv} and
  DELPHI \cite{Delphi_rpv} analyzed final states with multiple leptons
  and missing energy.
\item {\bf Six-Leptons} \\
  Multilepton final states can also arise from slepton decays in
  (R-parity conserving) gauge mediated SUSY breaking scenarios, where the missing energy is
  carried away by a gravitino.  The ALEPH search for 6 lepton final
  states is described in Ref. \cite{Aleph_gmsb6}.
\item {\bf Squark Searches} \\ 
Squark pair production at LEP can lead to a final state with two acoplanar jets, where each jet has small invariant mass, accompanied by missing energy (and possibly additional leptons).
The searches of OPAL \cite{Abbiendi:2002mp} and ALEPH \cite{Heister:2002hp} can pick up the signal when the Higgs decays to lepton jets carrying a large number of leptons, while the associated $Z$ decays invisibly.      
The OPAL search turns out to be especially constraining, due to the fact that the number of observed events is well below the expected SM background.  
    
\end{itemize}

\subsection{Tevatron Searches}
\label{sec:tevatron-searches}

The Tevatron experiments search for lepton jets in a noisier hadronic environment.  Even so, due to the large
Higgs production cross-section and the high luminosity,   discovery may be within reach with possibly many
light Higgs-to-lepton jets events already on tape.  
We identify the following relevant searches\footnote{Multilepton signatures were also addressed in Ref. \cite{cdfmm} where an excess of multi-muon events was reported. 
This search focuses on events with displaced vertices, and therefore it is not relevant for the Hidden Higgs signal we consider in this paper. 
In any case, cross sections required to address the CDF multi-muon excess are orders of magnitude larger than the Higgs production cross section at the Tevatron.} 
 
\begin{itemize}
\item{\bf Dark Photon Search}\\
  Recently, the D0 collaboration has made a search~\cite{hiddenjuri}
  for hidden photons produced in neutralino decays. In addition to the
  lepton jet, a requirement for an isolated (ordinary) photon is
  made.  The photon requirement reduces significantly the expected signal from Higgs-to-lepton jets decays,
  where the photon can come only from initial state radiation.
  In addition, the D0 search requires the lepton jet to have
  only two leptons which is uncommon in our scenario.  

\item{\bf  NMSSM Hidden Higgs} \\
  Ref. \cite{andy} targeted a Higgs boson decaying into $4\mu$ and
  $2\tau 2\mu$ final states via an intermediate pair of pseudoscalar
  singlets of the NMSSM.  The search focused on the case where the
  pseudoscalar is fairly light, so that the muon and tau pairs to
  which it decays are highly collimated.  Consequently,
  Ref. \cite{andy} looked for isolated 
  muons with close
  companion tracks.  This topology can readily arise in 
  Higgs-to-lepton jet decays as
  long as the hidden photon is heavy enough to decay into muon pairs.
\end{itemize}
 
Studies of multilepton final states have been routinely performed at
the Tevatron, mostly in the context of SUSY searches.
The SM processes are unlikely to produce 3 or more energetic and {\em
  isolated} leptons, therefore such topologies offer clean channels to
search for new physics.  Although these searches typically target
isolated high-$p_T$ leptons, it is conceivable that a subset of our
signal events may yield muons and electrons passing the selection
criteria.  The most interesting from our perspective are 
\begin{itemize}
\item{\bf Trilepton Searches} \\
  Of all trilepton searches 
  Ref. \cite{Cdf_3luni} is singled
  out because it is based on the largest data sample of $3.2$ fb$^{-1}$. Moreover, the cuts on lepton $p_T$ and on missing
  transverse energy are relatively soft.  However, the isolation
  requirements are quite severe. In particular, all objects in the
  analysis are required to be separated by $\Delta R > 0.4$, which decreases
  the sensitivity to the Higgs-to-lepton jets signal.  Another search in Ref. \cite{Cdf_3llow} focused on di-muon pairs
  accompanied by a third lepton with a very low $p_T$ threshold of $5$~GeV.  In this case lepton isolation is determined
  by calorimeter deposits: less then 10 percent of the lepton $p_T$
  should be detected in the $\Delta R = 0.4$ cone around the lepton.
\item{\bf Like-Sign Dilepton Searches} \\
Ref. \cite{Abulencia:2007rd} focused on events with two energetic electrons or muons of the same electric charge and large invariant mass.
Such a pair can arise in the signal when the two selected leptons come from separate lepton jets, or when a lepton in a lepton jet is paired with another lepton from W or Z decays.   
Again the sensitivity to our signal is reduced by the isolation requirements: the sum of the transverse energy within $\Delta R = 0.4$ around the leptons must be less than 10 percent of the lepton $p_T$. 
 
\end{itemize}

\section{Hiding the Higgs}
\label{sec:benchmark-models}
In this section we discuss the implications of the experimental searches, listed in Section~\ref{s.ec},
for viable models where the Higgs decays dominantly to lepton jets.  After introducing our 
methodology in Section~\ref{subsec:methodology}, we show in Section~\ref{subsec:ConstrainObservables}
that the set of observables  (listed in Section~\ref{sec:exper-observ}) are constrained by LEP and the Tevatron, to a particular region which can accommodate a  light Higgs decaying to lepton jets.
In Section~\ref{subsec:benchmarks} we then present concrete benchmark models that hide the Higgs.

\subsection{Methodology}
\label{subsec:methodology}
We are interested in how well the 
 LEP and Tevatron searches listed in 
Section \ref{s.ec} constrain  the Higgs decaying to lepton jets.  No
searches have explicitly placed limits on this signal, thus the limits must be inferred from
simulation.  
We simulate Higgs production and decays to lepton jets using 
Monte Carlo 
and evaluate the efficiency of the above searches by making the appropriate
cuts on the produced signal events.  We use
{\tt Madgraph} \cite{madgraph} to simulate the Higgs production and decay into
the hidden sector, {\tt BRIDGE} \cite{bridge} to simulate the hidden sector
cascades that populate lepton jets, {\tt Slowjet} 
\cite{slowjet} for event analysis, including kinematic cuts, jet
clustering, and lepton isolation.  We do not simulate hidden sector
showering, which can be important for $g_d^2/4 \pi \gtrsim 0.1$
\cite{itay,Meade:2009rb}.   
It is important to keep in
mind that, due to the collimated nature of lepton jets, some tracks may fail to reconstruct and some leptons may fail lepton identification.
While we work with the ideal situation where this is not the case, but to set reliable limits on scenarios where the Higgs decays to lepton jets, a 
more comprehensive study with full detector simulation is necessary.   Such a study is beyond the scope of this paper.

 \subsection{Constraints on Experimental Observables}
\label{subsec:ConstrainObservables}

We now discuss how the experimental searches constrain viable Higgs--to--lepton jets signatures.  We consider the observables:  event topology,
lepton multiplicity, lepton species, and missing energy listed in Section \ref{sec:exper-observ}. The discussion 
includes all three production channels discussed in Section~\ref{s.p}.

\begin{figure}[t]
\hbox{\includegraphics[width=0.47\textwidth]{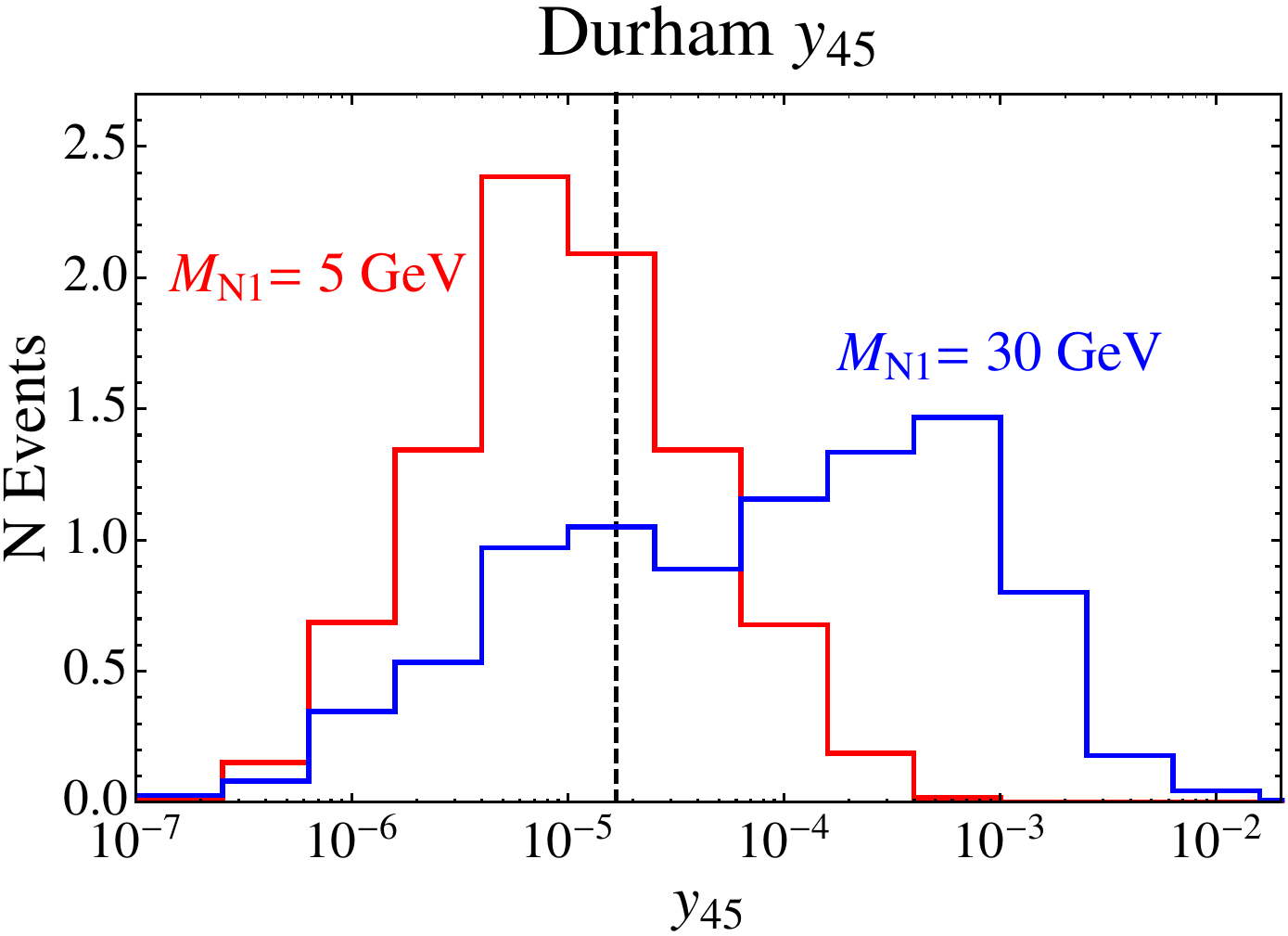} \hspace{0.02\textwidth} \includegraphics[width=0.47\textwidth]{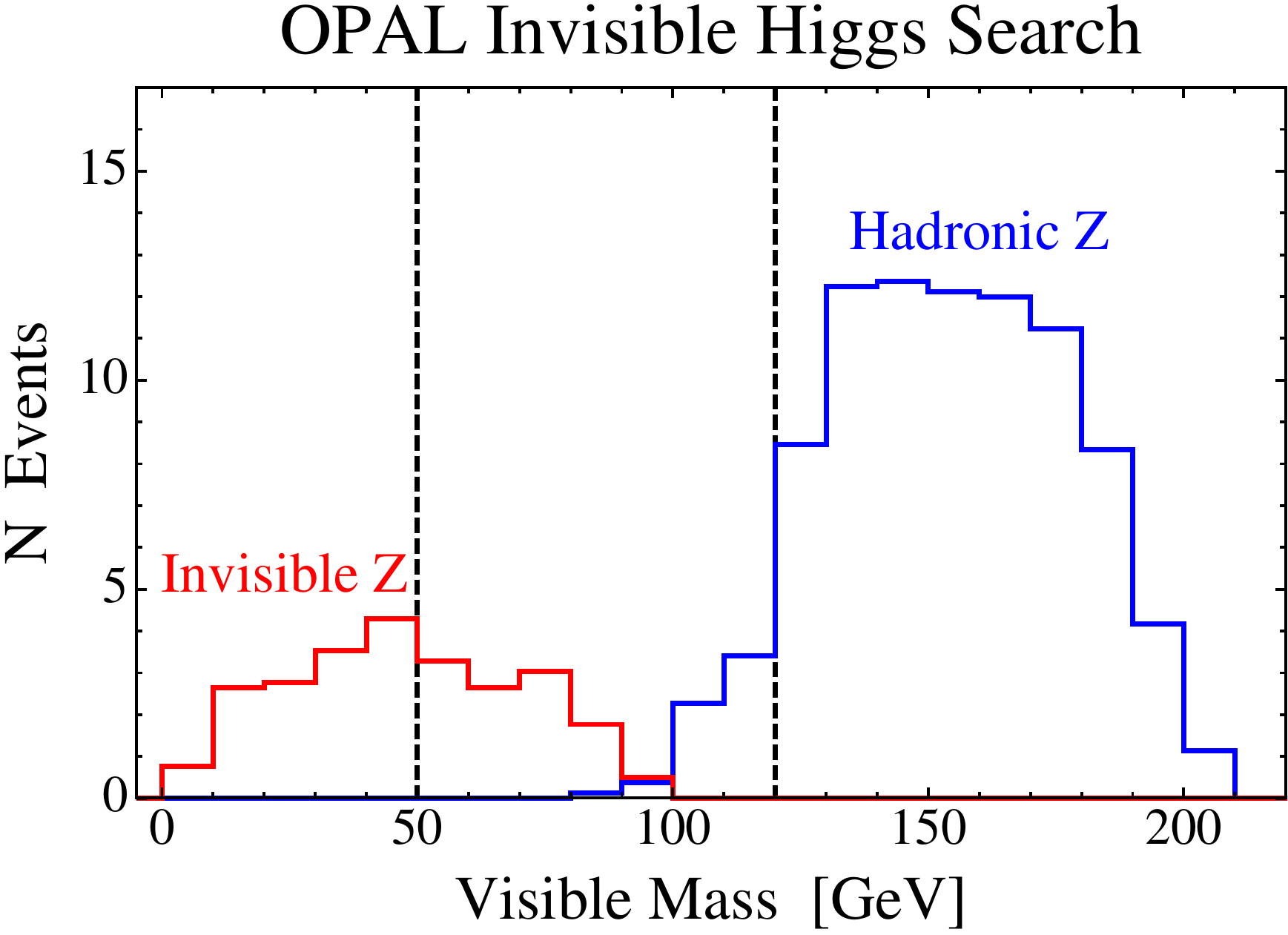} }
\caption{\small \setlength{\baselineskip}
{.9\baselineskip}
 {\it Left:} The Durham $y_{45}$ jet clustering parameter \cite{Moretti:1998qx} for Higgs-strahlung events with leptonic $Z$'s.  The ALEPH $h \rightarrow W W^*$ search \cite{Aleph_wwstar}, class {\bf 2c}, cuts on $y_{45} > 2 \times 10^{-5}$, selecting events with 5 well-separated objects ($y_{45}$ is small, if the event has less than 5 well-separated jets).  Due to  the leptonic $Z$, this search is less sensitive to models where the Higgs decays to two or less lepton jets.  {\it Right:} The visible mass distribution for Higgs-strahlung events with hadronic or invisible $Z$'s in the neutralino benchmark of Section~\ref{subsec:benchmarks}.  The OPAL $h \rightarrow E\!\!\!/_T$ search \cite{Opal_invisible} selects events with $50 \unit{GeV} < M_{\rm vis} < 120$ GeV\@.  Some, but not too much, missing energy per Higgs decay, $E_T \sim 50$ GeV, helps evade this search by keeping most invisible Z events below the window and hadronic (or leptonic) Z events above the window.  
}
\label{Fig:Durham45OpalInvis}
\end{figure}

We begin by arguing that several searches lead one to consider a two-jet topology for the lepton jets.  For instance, the neutralino channel is strongly constrained by the acoplanar jet search at LEP-1 \cite{Aleph_aco}, where
neutralino pairs can be produced in rare $Z$ decays.  While constraints on the Z-width allow the $Z$ branching
fraction to neutralinos to be as large as $10^{-3}$, the branching fraction to 3 or more lepton jets must be suppressed by $\sim
10^{-6}$ in order for the model not to be excluded.  Such low branching ratios are obtained for a very light neutralino,
$m_{\tilde N_1} \lesssim 5$ GeV, where the resulting event topology
consists of two back-to-back neutralino-jets (shown in the third panel of
Figure \ref{f.et}).  Decays with such a topology are not excluded because of the kinematical
similarity to the large hadronic $Z$ background.  A two-jet topology is also favored by the
$h\rightarrow W W^*$ search at LEP-2  \cite{Aleph_wwstar}.  
Especially constraining is a search
subclass consisting of a final state with two hard leptons, ($E_T > $
25 GeV and $E_T>$ 20 GeV), a softer lepton ($E_T > 8$ GeV), and at least two  additional
tracks.  This selection has a small SM background, which is
further reduced by using the Durham jet clustering algorithm to select
events with at least 5 well separated jet-like or single track objects.
This subclass is sensitive to our signal from $e^+e^-\to Zh$ production, if  the $Z$ decays leptonically (forming
two well-separated objects), while $h$ decays to three or more lepton jets.  Higgs production is therefore safe if the
Higgs decays to less than three lepton jets, as shown on the left panel of Figure
\ref{Fig:Durham45OpalInvis}.  We see that final states with two lepton jets are favored by
both LEP-1 and LEP-2.

We note, however, that the two-lepton jet topology is partially constrained by LEP-2.  In particular, searches for squark pair production target topologies with two QCD jets and missing energy, and these searches can be sensitive to Higgs decays to lepton jets accompanied by an invisible $Z$\@.  We find that the OPAL squark search,~\cite{Abbiendi:2002mp}, is most constraining, although because of the substantial SM background it does not have the sensitivity to exclude a 100 GeV Higgs decaying to lepton jets.  An exception is the sneutrino channel, which is particularly constrained by~\cite{Abbiendi:2002mp}, because of the extra lepton jet events due to direct sneutrino production through off-shell $Z$'s.   The ALEPH squark search,~\cite{Heister:2002hp}, is less sensitive to all channels due to tighter cuts that largely reduce the signal.  LEP-2 has also searched for Higgs decays to two QCD jets.  These searches do not seriously constrain Higgs decays to lepton jets because they require heavy flavor tagging and/or focus on Higgs decays without missing energy~\cite{Abbiendi:2003gd}.

The lepton multiplicity within lepton jets is also strongly
constrained by LEP-1 and LEP-2.  In particular, the neutralino cannot have a large branching fraction to invisible matter.  Indeed, at LEP-1, the monojet topology must be suppressed by $\sim 10^{-6}$, and
therefore the neutralino branching fraction to purely missing energy, by $\sim 10^{-3}$. 
The ALEPH search for $h \rightarrow 4 \tau$ \cite{kyle} at LEP-2 constrains the multiplicity further since the search is evaded by lepton jets
containing more than 4 leptons. 
 Meanwhile, the flavor of leptons
within lepton jets is most constrained at the Tevatron, where the D0
search for $h \rightarrow 4 \mu, \, 2 \mu \, 2 \tau$~\cite{andy} sets stringent
limits on final states containing muons with companion tracks.  Models
where lepton jets are electron only, $\mhid < 2 \, m_\mu$,
are unconstrained by this search.  When $\mhid > 2 \, m_\mu$, lepton jets consisting of exactly two muons must
be suppressed by $\sim 10^{-3}$, while lepton jets with more than two
leptons can spoil the track and calorimeter isolation requirements
placed on the muon pairs \cite{andy}.  We therefore find that high
multiplicity lepton jets are the least constrained, and are necessary for
models where lepton jets include muons.

The amount of missing energy per Higgs decay is most constrained by
the OPAL search for $h \rightarrow E\!\!\! /_T$ which selects events
with a visible mass in a wide window around the $Z$ mass, 50~GeV~$<
M_{\rm vis}<$~120~GeV\@.  Our signal can fall within this window when
Higgs decays include too much missing energy.  For a light Higgs, $m_Z
\sim m_h$, this search is also sensitive to the situation where the
$Z$ decays invisibly and the Higgs decays to lepton jets.  Then, some missing energy can lead to a signal with $M_{\rm vis} < 50$
GeV, evading the search.  We therefore find that hadronic and
invisible $Z$ decays constrain the missing energy from opposite
directions, and the optimal value is $E_T \sim 50$ GeV, as in the right panel of Figure
\ref{Fig:Durham45OpalInvis}.  We also find that the sneutrino channel is more difficult to
accommodate with this search than the other channels, because the
sneutrino decays produce neutrinos that carry substantial missing
energy.  We note that a hidden Higgs decaying to final states that include missing energy is also considered by Ref.~\cite{Chang:2007de}.

Finally, we comment that trilepton searches~\cite{Cdf_3luni,Cdf_3llow} and like-sign dilepton searches~\cite{Abulencia:2007rd}, are easily evaded by lepton jets due to the strong isolation requirements of these searches.  Consequently, most Tevatron searches do not constrain the Higgs-to-lepton jets scenario.  We summarize the consequences of existing searches at LEP and the
Tevatron for a light Higgs decaying to lepton jets,
\begin{itemize}
\item {\bf Two-Jet Topology:} The Higgs should decay to two lepton
  jets.  For the neutralino channel, $m_{\tilde N_1} \lesssim 10$ GeV.
\item {\bf High Lepton Multiplicity:} The lepton jets should have high
  lepton multiplicities, $\gtrsim 4$ leptons per lepton jet.
\item {\bf All Electron or Very High Multiplicity:} All-electron
  lepton jets, $\mhid < 2 \, m_\mu$, are the least
  constrained.  Very high multiplicity lepton jets can include muons,
  if the rate of events with isolated muon pairs, is suppressed by $\sim 10^{-3}$.
\item {\bf Some $\mathbf{E\!\!\! /_T}$:} The Higgs decays should
  produce some, but not too much, missing energy, $E\!\!\! /_T \sim
  50$ GeV.
\end{itemize}

\subsection{Benchmarks Models}
\label{subsec:benchmarks}

 We now present one benchmark model with a 100 GeV Higgs for each of the 
 decay channels: the neutralino channel (Section \ref{sec:neutralino-channel}), the sneutrino channel   (Section \ref{sec:sneutrino-channel}), and the singlet channel  (Section \ref{sec:singlet-channel}). 
 For the neutralino and
 sneutrino benchmarks the hidden photon decays only to electrons,
 whereas for the singlet benchmark muons are also produced.  The
 neutralino and singlet benchmarks pass all searches of sections \ref{sec:lep-1-searches}, \ref{sec:lep-2-searches}, and \ref{sec:tevatron-searches}
at the $2 \sigma$ level, hiding the Higgs from LEP and the Tevatron.
Efficiencies of the searches for each of the benchmarks are given in Appendix \ref{a.e}\@.
For the sneutrino model, we find some tension at accommodating all searches, when using the simple $U(1)_d$ hidden sector of Section \ref{sec:minimal-model}.  We present a benchmark model
 which is excluded by the ALEPH $h \rightarrow 4 \tau$ search~\cite{kyle} and the OPAL $2 j + E\!\!\!/_T$ squark search~\cite{Abbiendi:2002mp} at
 more than $3 \sigma$, but passes all the other searches within $2 \sigma$.  The tension is due to the extra lepton jet events following from $e^+e^-\to Z^*\to \tilde \nu \tilde \nu$, 
 the missing energy carried by the neutrinos produced
 in the sneutrino decays, and the difficulty
 at achieving a 2-jet structure with sufficient lepton multiplicity.  It is to be stressed that this latter difficulty is an
 artifact of the particular hidden sector model we are considering,
 and 
 can be ameliorated in models with longer hidden sector
 cascades or substantial showering.
 
The three benchmark models occur within the supersymmetric framework.  The constraints on these models are, in principle, sensitive to all soft parameters because SUSY partners may be produced at LEP and the Tevatron, and decay to lepton jets.  For this discussion, because were are interested specifically in the constraints on Higgs decays to lepton jets, we decouple all unrelated SUSY partners and only consider light soft parameters that play a role in Higgs decays.  For the neutralino channel, we specify the -ino spectrum, for the sneutrino channel we specify the left-handed slepton spectrum, and for the singlet channel we specify the parameters that determine the singlet VEV and $F$-term.  It would be interesting to relax this assumption, and study the constraints on a light Higgs accompanied by additional light SUSY partners decaying to lepton jets, and we leave this for future study. \\
 
\begin{figure}[t]
 \begin{center}
    \includegraphics[width=1.0\textwidth]{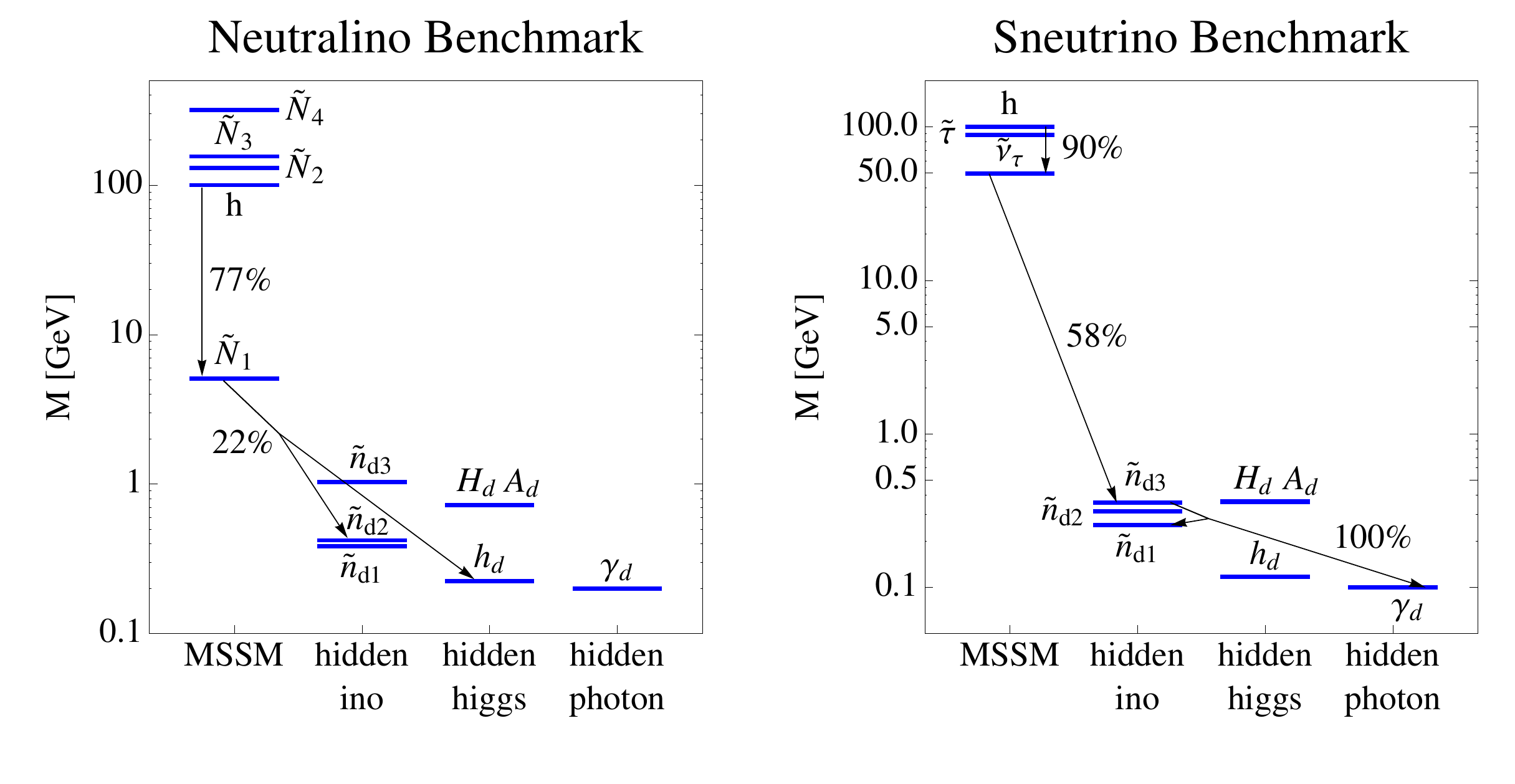}
     \end{center}
     \vspace{-1.cm}
\caption{\small\setlength{\baselineskip}
{.9\baselineskip} Spectra in the neutralino and sneutrino benchmark models. The dominant decay chain of the Higgs is denoted by arrows, labeled with branching ratios. Only the lowest lying states in the MSSM are shown. We also do not show the final decays into visible states in the last step of the decay chain. For the neutralino benchmark one has: $\tilde n_{d2}\to \tilde n_{d1} e^+e^-$, $h_d\to \gamma_d e^+ e^-$  where in both benchmarks $\gamma_d\to e^+e^-$  and $\tilde n_{d1}$ is stable.}
\label{f.spectra}
\end{figure}

\begin{figure}[t]
 \begin{center}
\includegraphics[width=0.5\textwidth]{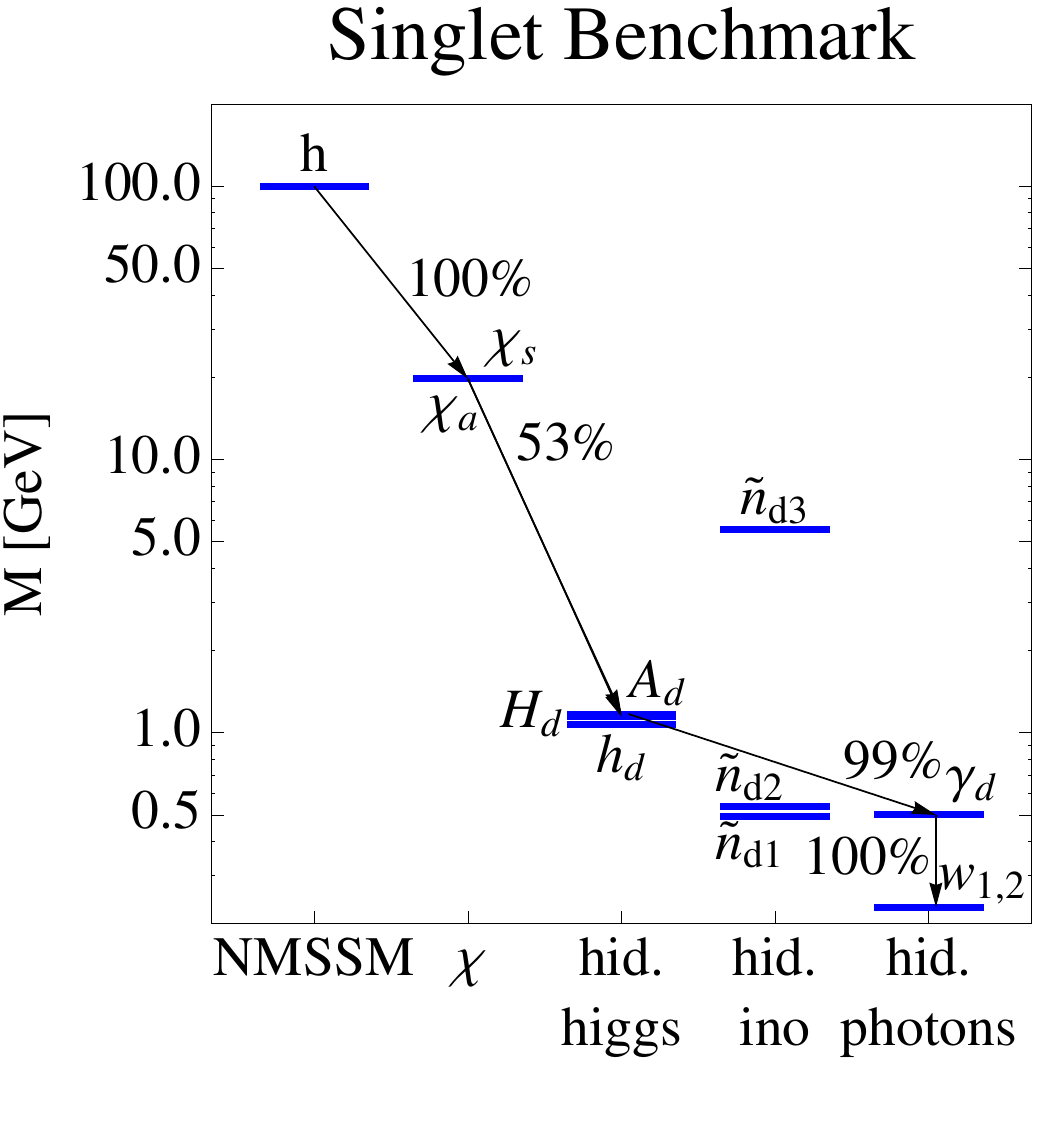}
     \end{center}
     \vspace{-1.cm}
\caption{\small \setlength{\baselineskip}
{.9\baselineskip} Spectrum of the singlet benchmark model. From the NMSSM states only the Higgs is shown, while other low lying states are also possible. The dominant decay chain of the Higgs is denoted by arrows, labeled with branching ratios. At the end of the decay chain $w_{\rm hid}^{1,2}$ decay to $e^+e^-$. The LSP is $\tilde n_{d1}$.} 
\label{f.spectra-singlet}
\end{figure}

\noindent
{\bf Neutralino Benchmark.}
For the neutralino benchmark we take $m_h = 100$ GeV, while the relevant
electroweak MSSM parameters are
\begin{equation}
\label{eq:NeutralinoParam1}
\begin{array} {lllll}
  \mu = 149 \unit{GeV},  &  M_1 = 13 \unit{GeV},& M_2 = 286 \unit{GeV}, &
  \tan\beta = 3.5, & \sin \alpha =  -0.28, 
\end{array}
\end{equation}
and the hidden sector parameters are,
\begin{equation}
\label{eq:NeutralinoParam2}
\begin{array} {llllll}
\mhid = 0.2  \unit{GeV},  & \muhid = 0.4 \unit{GeV}, &  \Mhid = 1
\unit{GeV},  & \tan \betahid = 4, & \sin \alpha_d = -0.27 \, , 
\end{array}
\end{equation}
where the hidden sector parameterization is defined in Appendix
\ref{a.abp}. The resulting spectrum is shown in the
left panel of Figure \ref{f.spectra}. 
Since $\mhid < 2 \, m_\mu$ the lepton jets are composed entirely of
electrons.    Note  that $m_{\tilde N_1}
= 5$ GeV, so that the Higgs decays produce boosted neutralinos.  The
resulting topology consists of 2 neutralino jets, avoiding the searches at LEP-1 and LEP-2.  The
lightest chargino has a mass of $m_{\tilde C_1} = 123$ GeV, while the second
lightest neutralino has a mass of $m_{\tilde N_2} = 130$ GeV\@.  These
masses are above the LEP-2 reach.  Also the Tevatron trilepton constraints
do not apply because $\tilde N_2$ dominantly decays to $\tilde N_1$
and an on-shell $Z$\@.  Trilepton searches veto events where opposite
sign leptons reconstruct an on-shell $Z$, in order to control the
electroweak background \cite{Cdf_3luni, Cdf_3llow}.  For this benchmark, the $Z$ decays to neutralinos with branching fraction $\Gamma_{Z \rightarrow 2 \tilde N_1} = 8 \times 10^{-4}$, which is consistent with the LEP-1 measurement of the $Z$ width \cite{Lep_zpole}. \\ 

\noindent
{\bf Sneutrino Benchmark.} For the sneutrino benchmark, we also take
$m_h \sim 100$ GeV. Here the Higgs decays into a pair of light tau
sneutrinos.  The relevant MSSM parameters are given by,
\begin{equation}
\label{eq:SneutrinoParam1}
\begin{array}{lll}
m_{L_3} = 77.3 \gev, & \tan \beta = 3.5, & \sin \alpha = -0.28,
\end{array}
\end{equation}
and the parameters that determine the hidden sector spectrum are,
\begin{equation}
\label{eq:SneutrinoParam2}
\begin{array}{lllll}
\mhid = 0.1  \gev,  & \muhid = 0.3 \gev, &  \Mhid = 0.3 \gev,  & \tan
\betahid = 4, & \sin \alpha_d= -0.27. 
\end{array}
\end{equation}
The spectrum is shown in the right panel of Figure \ref{f.spectra}.  
The hidden photon decays only to electrons.
The tau sneutrino has a mass of $m_{\tilde \nu_{\tau}} = 49.5$~GeV and 
the stau a mass of $m_{\tilde \tau} = 89$~GeV.  The stau and tau
sneutrino are pair produced at LEP through an off-shell $Z$, leading
to a lepton jet signal.  For reference, the LEP-2 
cross-sections of Higgs-strahlung, tau sneutrino pair production, and stau pair production
are, at $\sqrt s = 206$ GeV, $\sigma_{h Z}  \simeq0.3$ pb,
$\sigma_{\tilde \nu \tilde \nu} \simeq 0.3$ pb, and $\sigma_{\tilde
  \tau \tilde \tau} \simeq 0.1$ pb. 
  We find that this benchmark is ruled out by the ALEPH $h \rightarrow 4
\tau$ search~\cite{kyle} and the OPAL $2 j + E\!\!\!/_T$ squark search~\cite{Abbiendi:2002mp}, both at more than $3 \sigma$, while it passes all other searches at or below $2
\sigma$.  The ALEPH search is only sensitive to lepton jets with low
multiplicity and can be evaded by models that produce extra leptons through longer cascades or sufficient hidden sector showering. \\ 

 \noindent
{\bf Singlet Benchmark.}
The singlet benchmark also has a SM like Higgs at 100
GeV\@.  In this model, the NMSSM singlet $S$ couples to hidden sector
messengers $\chi$ and $\bar \chi$, which couple to the hidden sector Higgses $h_{1,2}$  (see equation \ref{eq:14}).  
The relevant NMSSM parameters \cite{martin}
are given by, 
\begin{equation}
\label{eq:SingletParam1}
\begin{array}{cccc}
\left<S \right> = 150\unit{GeV}, & \lambda =1, & \kappa=2, & \tan \beta = 5, \\
& A_\lambda = 10, & A_\kappa = 1 \, , & 
\end{array}
\end{equation}
and the hidden sector parameters are given by,
\begin{equation}
\label{eq:SingletParam2}
\begin{array}{cccc}
m_\chi = 160\unit{GeV},  & \muhid = 0.5\unit{GeV},& \mhid = 0.5 \unit{GeV}, & \tan \betahid = 2, \\
\Mhid =  5.5 \unit{GeV}, & \Bmuhid = - 0.5, \unit{GeV}^2, & \kappa_1 = 0.1, & \kappa_2 = 0.1.
\end{array}
\end{equation}
Here, $m_\chi$ denotes the mass of the fermionic messengers, which are
proportional to $\left< S \right>$.  The spectrum is shown in Figure \ref{f.spectra-singlet}.  The scalar messengers, $\chi_\pm$, are split
by $\left< F_S\right> $.  The lighter scalar messenger has a mass of
$20$ GeV\@, allowing for the Higgs decays, $h \rightarrow \chi_-
\chi_-^*$.  For this benchmark, we include muon production, which as
discussed above, requires substantial lepton multiplicity in order to
evade the D0 search for $h \rightarrow 4 \mu,\, 2 \mu \,2 \tau$ \cite{andy}.  We find
this difficult to achieve with the simplest $U(1)_d$ hidden sector, so
for this benchmark we extend the hidden sector to a larger non-Abelian
group, $SU(2)_d \supset U(1)_d$, where we take hidden sector cascades to
dominantly end with decays of the form $\gammahid \rightarrow w^1_d \, w_d^2 \rightarrow 2 \, l^- l^+$.

\section{Suggested Search Strategies}
\label{s.ss}

We have argued that signatures of the Higgs decaying into lepton jets
could have gone unnoticed by existing collider searches, especially if the leptons are collimated into two
jets, mimicking the topology of certain hadronic backgrounds.  Nevertheless, several
properties of the  lepton jets set them apart from QCD jets, and
dedicated searches  could very likely extract the signal from the SM
background.  Below we point out some features of the signal that could be
targeted by experiments in order to increase the sensitivity.  The
discussion is concise and qualitative; a more quantitative study is beyond
the scope of this paper and will be attempted elsewhere~\cite{lhc}.

\begin{itemize}

\item {\bf Hadronic energy deposition}.  An obvious consequence of the
  high lepton content of the jets is a small energy fraction deposited
  in the hadronic calorimeter.  Typical electrons stop in the
  electromagnetic calorimeter, while typical muons deposit of order
  1-2 GeV in the hadronic calorimeter.  Therefore experiments could
  search for jets with an anomalously small $E_{\rm had}/E_{\rm em}$ ratio.

\begin{figure}[t]
\hbox{
 \hspace{-1cm}
\includegraphics[width=1.08\textwidth]{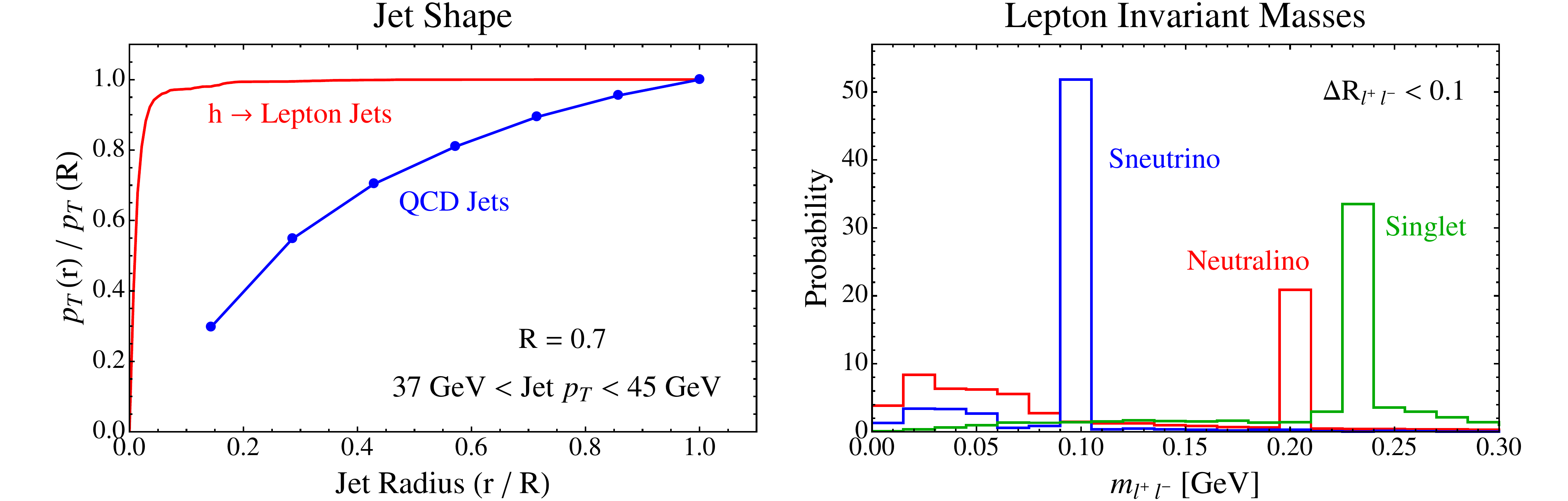}
}
\caption{\small \setlength{\baselineskip}
{.9\baselineskip} Lepton jet shape and constituent invariant masses.  {\it Left:}  We compare the shape of lepton jets produced by Higgs decays to the shape of QCD jets, as measured by CDF \cite{Acosta:2005ix}.  Jets of size $R = 0.7$ are identified by clustering the event with a midpoint-cone algorithm.  The jet shape is then defined as the $p_T$ fraction contained in smaller cones.  We see that lepton jets are much narrower than QCD jets.  {\it Right:} The invariant masses of all opposite sign lepton pairs, separated by $\Delta R < 0.1$, in Higgs decays to lepton jets.  We see clear invariant mass peaks corresponding to the hidden sector photon masses, despite the background of wrong-pair combinatorics.  Both jet shape and invariant mass peaks can be used to overcome the QCD background and discover lepton jets.
}
\label{f.im}
\end{figure}
   
\item {\bf Event shapes}. Since the leptons originate from highly
  boosted objects, lepton jets are slimmer than typical QCD jets.
  This is shown in \fref{im} where we compare the transverse momentum as
  a function of the jet radius for lepton jets and QCD
  jets~\cite{Acosta:2005ix}.  The main parameter setting the size
  (defined, e.g., by the cone where 90\% of the jet energy resides) is
  the ratio of the hidden photon mass to the electroweak scale.   For the 
  parameter choices
  considered in this paper, typical lepton jet sizes are $\Delta R
  \lesssim 0.1$, compared to $\Delta R \sim 0.7$ for QCD jets.  We stress
  that this estimates will change if hidden sector showering is
  significant, $g_d^2/4 \pi \gtrsim 0.1$.  An experiment could therefore impose cuts on the
  energy or $p_T$ in the $0.1 < \Delta R < 0.4$ cone, as proposed
  in~\cite{itay}.

\item {\bf Pair invariant mass}.  Many lepton pairs in the jets
  originate from two-body decays of the hidden photons.  Therefore the
  invariant mass of the pair reconstructs the hidden photon mass.  One
  can pair opposite sign same flavor leptons separated by, say,
  $\Delta R = 0.1$, and compute the invariant mass of all the pairs.  The signal sample displays a
 prominent peak at the invariant mass of the hidden photon
  despite the combinatoric background.   The peaks are clearly visible in \fref{im} for the three benchmark
  points described in Section~\ref{sec:benchmark-models}.

\item {\bf Leptons in jets}.  Lepton jets in our scenario are composed
  of electrons, and possibly muons, with high multiplicities of
  leptons within each jet.  QCD jets also contain leptons, but many of them originate from decays  of pions and kaons in-flight.  Prompt or almost prompt leptons can be
  produced by semileptonic decays of bottom and charm mesons, but these
  typically lead to 1 or 2 prompt leptons per jet.  Experiments could
  therefore search for jets with an anomalous lepton content.
  Repeating the analysis of~\cite{kyle} without demanding the small
  number of tracks inside the lepton jet, but using instead the jet shape,
  hadronic energy or lepton ID information to control the background, could
  allow for sufficient sensitivity to discover a light Higgs decaying to lepton jets.
 
 \end{itemize}

The above characteristics of lepton jets can be used to develop Tevatron searches, and 
to re-analyze old LEP-1 and LEP-2 data.   For
a 100 GeV hidden Higgs there are enough potential signal events to make such searches promising. We predict on the order of 100 Higgs events at LEP-2, and on the
order of $10^4$ Higgs events at the Tevatron.  Furthermore, the light neutralino scenario could lead to up to $10^4$ lepton
jet events at LEP-1.  This is substantially more than in the case
where no light SUSY state is present in which case the $Z$ branching fraction to the
hidden sector is suppressed by $\epsilon^2\lesssim 10^{-6}$ and the
number of lepton jets at LEP-1 is at most
$\mathcal{O}(1)$ \cite{baumgart}.  We conclude that a huge event sample might
be buried in the existing data waiting to be uncovered!  The precise
sensitivity of experiments crucially depends on the experimental
ability to reconstruct nearby tracks, untangle overlapping
calorimetric showers, and identify nearby leptons.  These issues are
difficult to estimate without a full and accurate detector simulation
and therefore input from the experimental community is vital.

\section{Conclusions and Outlook}
\label{s.c}

The mass and the decay width of a SM-like Higgs are strongly constrained
by LEP and Tevatron searches.  In the supersymmetric framework the
existing bounds imply the need for fine tuning.  These bounds (and
hence the fine tuning) may be ameliorated if the Higgs has non-standard
decays.  Similarly, existing bounds on the masses of SM superpartners
may be evaded if they decay unconventionally.  It is therefore
conceivable that both the Higgs and (possibly some of) the superpartners are
sufficiently light to have been copiously produced at LEP and
the Tevatron.  Searches for standard decays could have 
missed such
hidden states.

In this paper we studied the prospects for the Higgs, and possibly the
lightest neutralino or sneutrino, to be hidden at colliders due to
their dominant decays into states that are part of a low scale hidden
sector weakly coupled to the SM.  The hidden states then subsequently
cascade decay back to electrons and muons.  The low, ${\cal O}({\rm GeV})$,
mass gap in the hidden sector together with the hidden cascade decays,
imply a significant number of collimated final state leptons known as
lepton jets.  Here we have considered three channels where the Higgs first
decays to two light neutralinos, light sneutrinos or directly to the hidden sector
fields.  Due to the mixing between the SM and the hidden sector, the
lightest visible supersymmetric particle is no longer stable and
hidden cascade decays occur.  In all cases, the branching fraction of
the Higgs to the hidden sector can dominate.

 To test these scenarios we have identified the main experimental
 observables that characterize the collider signals of these models.   We then simulated and
 studied existing SUSY and Higgs searches at LEP and the Tevatron that are potentially
 sensitive to lepton jets.  Quite surprisingly, we found that the
 Higgs can evade detection if it {\it decays to some amount of missing
 energy together with many non-isolated electron or muons, all residing in two jet-like
 structures}.  The topology and the lepton multiplicity depend on the spectrum of
 both the hidden and the visible sectors.  Interestingly, the required
 phenomenology is easily produced by the minimal hidden
 sector model.
Our study suggests that  bounds on many additional SM superpartners
may be significantly weakened in the presence of a low scale hidden
sector.  Consequently, other superpartners, not studied in this work,
may be hidden at LEP or the Tevatron.  

 It would be very interesting to assess the potential for a discovery
 of this scenario in colliders~\cite{lhc}.  One very effective path
 would be to reanalyze the old LEP data.  A dedicated LEP-2 search for
$Z$+lepton jets should easily extract the signal of Higgs-to-lepton jet decays from the background or
 significantly improve the bound on the mass of the Higgs in this scenario.  Revisiting LEP-1 data may also be rewarding, because in the
 light neutralino scenario as many as $10^4$ lepton jets (which we dub
 neutralino jets) could have been produced.
Finally, the
prospects at the Tevatron and the LHC also look promising, given the large
number of final-state leptons.  
   All experiments could increase
their sensitivity to our scenario by zooming in on narrow jets with
small hadronic deposits, attempting reconstruction of prompt leptons
inside jets, counting their multiplicities, and studying the invariant
mass distribution of close lepton pairs.  
So far dedicated experimental searches were limited to finding
isolated lepton pairs.  Our work suggests
that this approach could be too narrow, and higher multiplicities of
collimated leptons should also be targeted.  To do so a serious study
incorporating detector simulation is therefore warranted.

Finally, we stress that the relevance of our work extends beyond this
particular hidden Higgs scenario.  Even if the Higgs boson is heavier
than $115$ GeV, Higgs and/or SM superpartners decaying to light SM
states via a hidden sector is a phenomenological possibility that
deserves attention.  The standard new physics signatures may be
altered or completely absent, and the signal would be missed unless it
is specifically searched for. Multiple theoretical possibilities
should be explored in order to ensure that such a scenario is not
overlooked at the LHC.

\section*{Acknowledgments} 
 We thank Nima Arkani-Hamed, Kyle Cranmer, Csaba Csaki, JiJi Fan, David Krohn, Matt Reece, Liantao Wang and Itay Yavin
 for useful discussions. ÊWe would especially like to thank Christophe Delaere, Yuri Gershtein, and Chris Tully for very useful discussions about LEP and the Tevatron. ÊThe work of AF was supported in part by
 the Department of Energy grant DE-FG02-96ER40949. ÊJTR is
 supported by an NSF graduate fellowship. ÊTV is supported by DOE
 grant DE-FG02-90ER40542. The work of JZ is supported in part by 
the European Commission RTN network, Contract 
No. MRTN-CT-2006-035482 (FLAVIAnet) and by the 
Slovenian Research Agency.

\appendix
\section{Notation}
\label{a.abp}
The visible sector is the ordinary MSSM, and we follow the notation of \cite{martin}.
The hidden sector is a broken supersymmetric U(1) gauge theory with 2 Higgs multiplets $h_{1,2}$ of opposite U(1) charge and the superpotential
$W = \mu_d h_1 h_2$ \cite{itay}.
Such a theory is completely specified by 3 scalar soft mass terms $m_{1,2}^2$, $B \mu_d$, the hidden bino mass $M_d$, the mu-term $\mu_d$ and the gauge coupling $g_d$.
We do not impose any constraints on these parameters, in particular we do not assume any specific mechanism of supersymmetry breaking mediation to the dark sector.
The scalar potential takes the form
\beq
V = (m_1^2 + \mu_d^2)|h_1|^2 + (m_2^2 + \mu_d^2)|h_2|^2 + (B \mu_d h_1 h_2 + \hc) + {g_d^2 \over 2} \left (|h_1|^2 - |h_2|^2 \right )^2.  \eeq
We are interested in the regions of the parameter space where the hidden scalars acquire vevs, $\la h_1\ra = \sin \beta_d v_d$, $\la h_2 \ra = \cos \beta_d v_d$, which gives the hidden photon the mass $m_{\gamma D} = g_d v_d$.
In the broken phase, it is convenient to trade the scalar soft mass terms for more physical parameters: $v_d$, $\beta_d$ and the mixing angle $\alpha_d$ of the two CP-even mass eigenstates $h_d,H_d$.
The physical scalars are embedded in the Higgs fields as  \bea
\label{e.dhp}
h_1 &=& {\sin \beta_d v_d + \cos \alpha_d h_d + \sin \alpha_d H_d  + i \cos \beta_d   A_d  
\over \sqrt 2}
\nn  h_2 &=& {\cos \beta_d  v_d -  \sin \alpha_d h_d  + \cos \alpha_d H_d  + i \sin \beta_d  A_d  
\over \sqrt 2}
\eea
where  $A_d$ is the CP-odd scalar eigenstate.
At the tree level the mass of the lighter Higgs $h_d$ is constrained to be $\leq m_{\gamma D}$, in analogy with the MSSM.  However loop corrections from the hidden gauge and Higgs multiplets or from additional multiplets in the hidden sector can change that relation.  Additional couplings to heavier hidden states can also contribute to the hidden Higgs mass.
Therefore we take $m_{h_d}$ as a free parameter of order $m_{\gamma D}$ when constructing our benchmark models.   
In the fermionic sector the mass terms after U(1) breaking are \beq
- m_{\gamma D} \sin \beta_d  \ti h_1 \ti b  +   m_{\gamma D} \cos \beta_d  \ti h_2 \ti b   - {M_d \over 2} \ti b \ti b   - \mu_d  \ti h_1 \ti h_2 + \hc.  \eeq
The physical fermions are 3 dark neutralinos who are mixtures of hidden binos and higgsinos.  

\section{Efficiencies of the Searches}
\label{a.e}

\begin{table}[h!]
\bc
\begin{tabular}{c|c|ccccccc}
\hline\hline
 \multicolumn{8}{c}{LEP-1 searches}\\ \hline
 Search & Ref. & Obs. & Bckg.  & Neutr. & Sneutr.  & Singlet & Max. \\ \hline
Monojets & \cite{Aleph_mono}& 3 & 2.8 & $< 1$ & 0 & 0  & 6.6 
\\ \hline
Acoplanar & \cite{Aleph_aco}& 0 & 0.2 & $< 1$ & 0 & 0 & 3.8 
\\ \hline\hline
\multicolumn{8}{c}{LEP-2 searches}\\ \hline
 Search & Ref. & Obs. & Bckg.  & Neutr. & Sneutr.  & Singlet & Max.\\ \hline
$H \to 4 \tau$ & \cite{kyle}& 2 & 5.09 &  1 & {\bf 15} & 1 & 5.0 
\\ \hline
$H \to E\!\!\!\!/$ & \cite{Opal_invisible} & 8 & 11&  2 & 5 & 3  & 7.5 
\\ \hline
$H \to WW^*$2c & \cite{Aleph_wwstar} & 0 & 0.3 & 2 & $<1$ & 2  & 3.8 
\\ \hline
$H \to WW^*$2t & \cite{Aleph_wwstar} & 1 & 1.2 & 1 & 1 & 3  & 5.0 
\\ \hline
6l  & \cite{Aleph_gmsb6}& 1 & 1.1  & $<1$ & 4 & $<1$ & 5.0 
\\ \hline
$2j+ E\!\!\!\!/$ (OPAL) &\cite{Abbiendi:2002mp} & 13 & 19.8  &  8 & {\bf 35} &  7  & 7.8   
\\ \hline
$2j+ E\!\!\!\!/$ (ALEPH) &\cite{Heister:2002hp}& 19 & 15.9  & 7 & 3 &  1  & 14.5   
\\ \hline
$2j+2l+ E\!\!\!\!/$ &\cite{Heister:2002hp}& 5 & 3  & 2 & 4 &  5  & 9.0   
 \\ \hline\hline
 \multicolumn{8}{c}{Tevatron searches}\\ \hline
 Search & Ref. & Obs. & Bckg.  & Neutr. & Sneutr.  & Singlet & Max. \\ \hline
Dark photon  & \cite{hiddenjuri} & 7 & 8  & $\sim 1$ & $<1$ &  $<1$  & 7.9   
\\ \hline
$H \to 4 \mu$  & \cite{andy}& 2 & 2.2  & 0 & 0 &  2  & 5.8   
\\ \hline
Unified 3l  & \cite{Cdf_3luni}& 1 & 1.47 &  $<1$ & $<1$ & $<1$  &  3.7 
 \\ \hline
Low $p_T$ 3l  & \cite{Cdf_3llow}& 1 & 0.4 &  $<1$ & $<1$ & $<1$  &  5.4 
\\ \hline
Like-sign 2l  & \cite{Abulencia:2007rd} & 13 & 7.8 & 1 & $<1$ &  $<1$  & 14.7  
\\\hline\hline
\end{tabular}
\caption{A compilation of relevant searches for constraining the Higgs-to-lepton jet events.}\label{tab.searches}
\ec 
\end{table}

In this appendix we quote the efficiencies of selected experimental analyses to the signals from the 3 benchmark models of Section \ref {s.p}. In Table \ref{tab.searches} we list the number of the  observed events in each of the searches after all the cuts are applied (Obs.),  the expected number of  SM background events (Bckg.), and the predicted number of signal events for the three benchmarks: the neutralino channel (Neutr.), the sneutrino channel (Sneutr.) and the singlet channel (Singlet).
This should be compared to the maximum number of signal events (Max.) allowed at the 97.7\% confidence level. This corresponds to a 2-$\sigma$ one sided "exclusion" for Gaussian errors.  
The confidence level is determined using the $CL_s$ prescription used at LEP for Higgs searches and takes into account downward fluctuations in the data compared to the expected background \cite{Read:CLs,Junk:1999kv}. The searches that do not pass the 2-sigma threshold are denoted in bold.

The numbers of observed events and the background estimates given in Table \ref{tab.searches} refer to the following: \bi
\item $H \to 4 \tau$ search: invisible Z channel, $n_{1,2}^{\rm track} =$2 or 4, see Table 3 of \cite{kyle},
\item Invisible Higgs search in OPAL:  more-than-2-jet events with $M_{miss}$ in the 100-104 GeV bin, see Fig.4 in \cite{Opal_invisible},
\item  $H \to WW^*$ search in ALEPH: the class 2c and 2t  defined in Table 3 of \cite{Aleph_wwstar},
\item   Six-lepton  search in ALEPH:  small $\Delta m$ selection defined in Table 3.4 of \cite{Aleph_gmsb6},
\item $2j+ E\!\!\!\!/$ search in OPAL and ALEPH: selection A, see Table 2  of \cite{Abbiendi:2002mp},  and large $\Delta m$ AJ selection (years 1999-2000), see Table 5 of  \cite{Heister:2002hp}, 
\item $2j+2l+ E\!\!\!\!/$ in ALEPH: to the large $\Delta m$ selection (years 1999-2000), see Table 5 of \cite{Heister:2002hp}
\item Dark photon search in D0: electron pairs  with 0.2-0.4 GeV invariant mass, see Fig. 2 of \cite{hiddenjuri}, 
\item Unified trilepton search in CDF: trileptons defined by the (least tight) ttt cut, see Table 1 of \cite{Cdf_3luni},
\item Like-sign dilepton search in CDF: dilepton events with missing energy, see Table 2 of \cite{Abulencia:2007rd}.
\ei  

The signal predictions were obtained using monte carlo as explained in Section \ref{subsec:methodology}.



\begin{thebibliography}{nn}

\bibitem{Barate:2003sz}
  R.~Barate {\it et al.}  
  Phys.\ Lett.\  B {\bf 565}, 61 (2003)
  [arXiv:hep-ex/0306033].

\bibitem{gfitter}
  H.~Flacher, M.~Goebel, J.~Haller, A.~Hocker, K.~Moenig and J.~Stelzer,
  Eur.\ Phys.\ J.\  C {\bf 60}, 543 (2009)
  [arXiv:0811.0009 [hep-ph]].
  
\bibitem{O'Connell:2006wi}
  R.~E.~Shrock and M.~Suzuki,
  Phys.\ Lett.\  B {\bf 110}, 250 (1982).
  B.~A.~Dobrescu and K.~T.~Matchev,
  JHEP {\bf 0009}, 031 (2000)
  [arXiv:hep-ph/0008192];
  D.~O'Connell, M.~J.~Ramsey-Musolf and M.~B.~Wise,
 Phys.\ Rev.\ ÊD {\bf 75}, 037701 (2007)
 [arXiv:hep-ph/0611014].
    O.~Bahat-Treidel, Y.~Grossman and Y.~Rozen,
  JHEP {\bf 0705}, 022 (2007)
  [arXiv:hep-ph/0611162];
K.~Cheung, J.~Song and Q.~S.~Yan,
  Phys.\ Rev.\ Lett.\  {\bf 99}, 031801 (2007)
  [arXiv:hep-ph/0703149].
  
  
\bibitem{Lep_4b}
  S.~Schael {\it et al.}  
  Eur.\ Phys.\ J.\  C {\bf 47}, 547 (2006)
  [arXiv:hep-ex/0602042].

\bibitem{Chang:2008cw}
  S.~Chang, R.~Dermisek, J.~F.~Gunion and N.~Weiner,
  Ann.\ Rev.\ Nucl.\ Part.\ Sci.\  {\bf 58}, 75 (2008)
  [arXiv:0801.4554 [hep-ph]].
  
\bibitem{Dermisek:2009si}
  R.~Dermisek,
  Mod.\ Phys.\ Lett.\  A {\bf 24}, 1631 (2009)
  [arXiv:0907.0297 [hep-ph]].

  
  \bibitem{DG1}
  R.~Dermisek and J.~F.~Gunion,
  Phys.\ Rev.\ Lett.\  {\bf 95}, 041801 (2005)
  [arXiv:hep-ph/0502105];
 R.~Dermisek and J.~F.~Gunion,
  Phys.\ Rev.\  D {\bf 75}, 075019 (2007)
  [arXiv:hep-ph/0611142].
    R.~Dermisek and J.~F.~Gunion,
  arXiv:1002.1971 [hep-ph].

\bibitem{CFW}
 S.~Chang, P.~J.~Fox and N.~Weiner,
  JHEP {\bf 0608}, 068 (2006)
  [arXiv:hep-ph/0511250].

\bibitem{buried}
  B.~Bellazzini, C.~Csaki, A.~Falkowski and A.~Weiler,
  Phys.\ Rev.\  D {\bf 80}, 075008 (2009)
  [arXiv:0906.3026 [hep-ph]];
  B.~Bellazzini, C.~Csaki, A.~Falkowski and A.~Weiler,
  arXiv:0910.3210 [hep-ph].

  
\bibitem{linda}
  L.~M.~Carpenter, D.~E.~Kaplan and E.~J.~Rhee,
  Phys.\ Rev.\ Lett.\  {\bf 99}, 211801 (2007)
  [arXiv:hep-ph/0607204].

\bibitem{cpv}
M.~S.~Carena, J.~R.~Ellis, A.~Pilaftsis and C.~E.~M.~Wagner,
  Nucl.\ Phys.\  B {\bf 586}, 92 (2000)
  [arXiv:hep-ph/0003180].
  M.~S.~Carena, J.~R.~Ellis, S.~Mrenna, A.~Pilaftsis and C.~E.~M.~Wagner,
  Nucl.\ Phys.\  B {\bf 659}, 145 (2003)
  [arXiv:hep-ph/0211467].


\bibitem{kyle}
  ALEPH~Collaboration,
  arXiv:1003.0705 [hep-ex].

  
  \bibitem{Opal_general}
  G.~Abbiendi {\it et al.}  [OPAL Collaboration],
  Eur.\ Phys.\ J.\  C {\bf 27}, 311 (2003)
  [arXiv:hep-ex/0206022].


\bibitem{nn} 
  N.~Arkani-Hamed and N.~Weiner,
  JHEP {\bf 0812}, 104 (2008)
  [arXiv:0810.0714 [hep-ph]].
  
\bibitem{itay}
  C.~Cheung, J.~T.~Ruderman, L.~T.~Wang and I.~Yavin,
  arXiv:0909.0290 [hep-ph].
  
\bibitem{baumgart}
  M.~Baumgart, C.~Cheung, J.~T.~Ruderman, L.~T.~Wang and I.~Yavin,
  JHEP {\bf 0904}, 014 (2009)
  [arXiv:0901.0283 [hep-ph]].


\bibitem{kathryn}
 M.~J.~Strassler and K.~M.~Zurek,
 Phys.\ Lett.\ ÊB {\bf 651}, 374 (2007)
 [arXiv:hep-ph/0604261].
 T.~Han, Z.~Si, K.~M.~Zurek and M.~J.~Strassler,
 JHEP {\bf 0807}, 008 (2008)
 [arXiv:0712.2041 [hep-ph]].
M.~J.~Strassler,
 arXiv:hep-ph/0607160.

\bibitem{pamela}
  O.~Adriani {\it et al.}  [PAMELA Collaboration],
  Nature {\bf 458}, 607 (2009)
  [arXiv:0810.4995 [astro-ph]].
  
  
  
\bibitem{Collaboration:2008aaa}
  F.~Aharonian {\it et al.}  [H.E.S.S. Collaboration],
  Phys.\ Rev.\ Lett.\  {\bf 101}, 261104 (2008)
  [arXiv:0811.3894 [astro-ph]];
  Astron.\ Astrophys.\  {\bf 508}, 561 (2009)
  [arXiv:0905.0105 [astro-ph.HE]].

\bibitem{Abdo:2009zk}
  A.~A.~Abdo {\it et al.}  [Fermi LAT Collaboration],
  Phys.\ Rev.\ Lett.\  {\bf 102}, 181101 (2009)
  [arXiv:0905.0025 [astro-ph.HE]].

\bibitem{tracy}
  N.~Arkani-Hamed, D.~P.~Finkbeiner, T.~R.~Slatyer and N.~Weiner,
  Phys.\ Rev.\  D {\bf 79}, 015014 (2009)
  [arXiv:0810.0713 [hep-ph]].

\bibitem{MPSV}
  P.~Meade, M.~Papucci, A.~Strumia and T.~Volansky,
  arXiv:0905.0480 [hep-ph].
  
\bibitem{Pospelov:2008jd}
  M.~Pospelov and A.~Ritz,
  Phys.\ Lett.\  B {\bf 671}, 391 (2009)
  [arXiv:0810.1502 [hep-ph]].


\bibitem{Rothstein:2009pm}
  I.~Z.~Rothstein, T.~Schwetz and J.~Zupan,
  JCAP {\bf 0907}, 018 (2009)
  [arXiv:0903.3116 [astro-ph.HE]].
  
\bibitem{Cheung:2009qd}
  C.~Cheung, J.~T.~Ruderman, L.~T.~Wang and I.~Yavin,
  Phys.\ Rev.\  D {\bf 80}, 035008 (2009)
  [arXiv:0902.3246 [hep-ph]];
  A.~Katz and R.~Sundrum,
  JHEP {\bf 0906}, 003 (2009)
  [arXiv:0902.3271 [hep-ph]].
  
\bibitem{Chun:2008by}
E.~J.~Chun and J.~C.~Park,
JCAP {\bf 0902} (2009) 026
[arXiv:0812.0308 [hep-ph]].


\bibitem{josh} 
J.~T.~Ruderman and T.~Volansky,
  arXiv:0908.1570 [hep-ph];
X.~Chen,
JCAP {\bf 0909} (2009) 029
[arXiv:0902.0008 [hep-ph]];
J.~Mardon, Y.~Nomura and J.~Thaler,
Phys.\ Rev.\  D {\bf 80} (2009) 035013
[arXiv:0905.3749 [hep-ph]].
  
 
 \bibitem{Strassler:2006ri}
M.~J.~Strassler and K.~M.~Zurek,
Phys.\ Lett.\  B {\bf 661} (2008) 263
[arXiv:hep-ph/0605193].
  
  
\bibitem{james}
  S.~Gopalakrishna, S.~Jung and J.~D.~Wells,
  Phys.\ Rev.\  D {\bf 78}, 055002 (2008)
  [arXiv:0801.3456 [hep-ph]].
  
  \bibitem{Batell:2009di}
B.~Batell, M.~Pospelov and A.~Ritz,
Phys.\ Rev.\  D {\bf 80} (2009) 095024
[arXiv:0906.5614 [hep-ph]].

\bibitem{Holdom:1985ag}
B.~Holdom,
Phys.\ Lett.\  B {\bf 166} (1986) 196.

\bibitem{Pospelov:2007mp}
M.~Pospelov, A.~Ritz and M.~B.~Voloshin,
Phys.\ Lett.\  B {\bf 662} (2008) 53
[arXiv:0711.4866 [hep-ph]].

\bibitem{Batell:2009yf}
B.~Batell, M.~Pospelov and A.~Ritz,
Phys.\ Rev.\  D {\bf 79} (2009) 115008
[arXiv:0903.0363 [hep-ph]].

\bibitem{Essig:2009nc} 
M.~Ahlers, J.~Jaeckel, J.~Redondo and A.~Ringwald,
  Phys.\ Rev.\  D {\bf 78}, 075005 (2008)
  [arXiv:0807.4143 [hep-ph]];
 Y.~Bai and Z.~Han,
  Phys.\ Rev.\ Lett.\  {\bf 103}, 051801 (2009)
  [arXiv:0902.0006 [hep-ph]];
R.~Essig, P.~Schuster and N.~Toro,
Phys.\ Rev.\  D {\bf 80} (2009) 015003
[arXiv:0903.3941 [hep-ph]];
M.~Reece and L.~T.~Wang,
JHEP {\bf 0907} (2009) 051
[arXiv:0904.1743 [hep-ph]];
J.~D.~Bjorken, R.~Essig, P.~Schuster and N.~Toro,
Phys.\ Rev.\  D {\bf 80} (2009) 075018
[arXiv:0906.0580 [hep-ph]];
 M.~Goodsell, J.~Jaeckel, J.~Redondo and A.~Ringwald,
  JHEP {\bf 0911}, 027 (2009)
  [arXiv:0909.0515 [hep-ph]];
 M.~Freytsis, G.~Ovanesyan and J.~Thaler,
arXiv:0909.2862 [hep-ph];
P.~Schuster, N.~Toro and I.~Yavin,
Phys.\ Rev.\  D {\bf 81} (2010) 016002
[arXiv:0910.1602 [hep-ph]];
 P.~Meade, S.~Nussinov, M.~Papucci and T.~Volansky,
arXiv:0910.4160 [hep-ph];
R.~Essig, P.~Schuster, N.~Toro and B.~Wojtsekhowski,
arXiv:1001.2557 [hep-ph].

\bibitem{MPZ}
 D.~E.~Morrissey, D.~Poland and K.~M.~Zurek,
 JHEP {\bf 0907}, 050 (2009)
 [arXiv:0904.2567 [hep-ph]].

\bibitem{Opal_invisible}
  G.~Abbiendi {\it et al.}  [OPAL Collaboration],
  arXiv:0707.0373 [hep-ex].
  
  \bibitem{Mason:2009qh}
J.~D.~Mason, D.~E.~Morrissey and D.~Poland,
Phys.\ Rev.\  D {\bf 80} (2009) 115015
[arXiv:0909.3523 [hep-ph]].

  
\bibitem{Djouadi:1997yw}
  A.~Djouadi, J.~Kalinowski and M.~Spira,
  Comput.\ Phys.\ Commun.\  {\bf 108}, 56 (1998)
  [arXiv:hep-ph/9704448].
  
   \bibitem{Meade:2009rb}
P.~Meade, M.~Papucci and T.~Volansky,
JHEP {\bf 0912} (2009) 052
[arXiv:0901.2925 [hep-ph]].
  
\bibitem{Amsler:2008zzb}
  C.~Amsler {\it et al.}  [Particle Data Group],
  Phys.\ Lett.\  B {\bf 667}, 1 (2008).
  


\bibitem{Aleph_aco}
  D.~Buskulic {\it et al.}  [ALEPH Collaboration],
  Phys.\ Lett.\  B {\bf 313}, 299 (1993).
  
  \bibitem{Aleph_mono}
  D.~Buskulic {\it et al.}  [ALEPH Collaboration],
  Phys.\ Lett.\  B {\bf 334}, 244 (1994).

\bibitem{Abulencia:2007rd}
  A.~Abulencia {\it et al.}  [CDF Collaboration],
  Phys.\ Rev.\ Lett.\  {\bf 98}, 221803 (2007)
  [arXiv:hep-ex/0702051].

\bibitem{Cdf_3luni}
The CDF Collboration, CDF Public Note 9817 (2009), {\it ''Update of the Unified Trilepton Search with 3.2 fb$^{-1}$ of Data,} available on www-cdf.fnal.gov/physics/exotic/r2a/20090521.trilepton\_3fb

\bibitem{Cdf_3llow}
   T.~Aaltonen {\it et al.}  [CDF Collaboration],
  Phys.\ Rev.\  D {\bf 79}, 052004 (2009)
  [arXiv:0810.3522 [hep-ex]].
  
\bibitem{Graesser:2007yj}
  M.~L.~Graesser,
  Phys.\ Rev.\  D {\bf 76}, 075006 (2007)
  [arXiv:0704.0438 [hep-ph]];
  M.~L.~Graesser,
  arXiv:0705.2190 [hep-ph].
  
  \bibitem{VistaSleuth}
 G.~Choudalakis, {et al.}, [CDF Collaboration], CDF note 8763, available at www-cdf.fnal.gov/physics/exotic/r2a/20070426.vista\_sleuth/publicPage.html
  
\bibitem{Lep_smHiggs}
  R.~Barate {\it et al.}  [LEP Working Group for Higgs boson searches],
  Phys.\ Lett.\  B {\bf 565}, 61 (2003)
  [arXiv:hep-ex/0306033].

\bibitem{Lep_invisible} 
    [LEP Higgs Working Group for Higgs boson searches],
  arXiv:hep-ex/0107032.
  
  \bibitem{Lep_zpole}   
[LEP], 
  Phys.\ Rept.\  {\bf 427}, 257 (2006)
  [arXiv:hep-ex/0509008].

\bibitem{Abbiendi:2003gd}
  G.~Abbiendi {\it et al.}  [OPAL Collaboration],
  Phys.\ Lett.\  B {\bf 597}, 11 (2004)
  [arXiv:hep-ex/0312042].

\bibitem{Aleph_wwstar}
  S.~Schael {\it et al.}  [ALEPH Collaboration],
  Eur.\ Phys.\ J.\  C {\bf 49}, 439 (2007)
  [arXiv:hep-ex/0605079].
  
  \bibitem{Aleph_rpv}
 A.~Heister {\it et al.}  [ALEPH Collaboration],
  Eur.\ Phys.\ J.\  C {\bf 25}, 1 (2002)
  [arXiv:hep-ex/0201013].

\bibitem{Delphi_rpv}
  J.~Abdallah {\it et al.}  [DELPHI Collaboration],
  Eur.\ Phys.\ J.\  C {\bf 36}, 1 (2004)
  [Eur.\ Phys.\ J.\  C {\bf 37}, 129 (2004)]
  [arXiv:hep-ex/0406009].
  
  \bibitem{Aleph_gmsb6}
  A.~Garcia-Bellido,
  arXiv:hep-ex/0212024.


   \bibitem{Abbiendi:2002mp}
G.~Abbiendi {\it et al.}  [OPAL Collaboration],
Phys.\ Lett.\  B {\bf 545} (2002) 272
[Erratum-ibid.\  B {\bf 548} (2002) 258]
[arXiv:hep-ex/0209026].

  \bibitem{Heister:2002hp}
A.~Heister {\it et al.}  [ALEPH Collaboration],
Phys.\ Lett.\  B {\bf 537} (2002) 5
[arXiv:hep-ex/0204036].

\bibitem{cdfmm}
  T.~Aaltonen {\it et al.}  [CDF Collaboration],
  arXiv:0810.5357 [hep-ex].


\bibitem{hiddenjuri}
  V.~M.~Abazov {\it et al.}  [D0 Collaboration],
  Phys.\ Rev.\ Lett.\  {\bf 103}, 081802 (2009)
  [arXiv:0905.1478 [hep-ex]].
  
  \bibitem{andy}
  V.~M.~Abazov {\it et al.}  [D0 Collaboration],
  Phys.\ Rev.\ Lett.\  {\bf 103}, 061801 (2009)
  [arXiv:0905.3381 [hep-ex]].

\bibitem{madgraph}
  J.~Alwall {\it et al.},
  JHEP {\bf 0709}, 028 (2007)
  [arXiv:0706.2334 [hep-ph]].

\bibitem{bridge}
  P.~Meade and M.~Reece,
  arXiv:hep-ph/0703031.

\bibitem{slowjet} {\tt SlowJet} is a private {\tt Mathematica} package for event
analysis, to be distributed upon demand.  

\bibitem{Moretti:1998qx}
  S.~Moretti, L.~Lonnblad and T.~Sjostrand,
  JHEP {\bf 9808}, 001 (1998)
  [arXiv:hep-ph/9804296].
 

\bibitem{Chang:2007de}
  S.~Chang and N.~Weiner,
  JHEP {\bf 0805}, 074 (2008)
  [arXiv:0710.4591 [hep-ph]].

\bibitem{martin}
  S.~P.~Martin,
  arXiv:hep-ph/9709356.

\bibitem{lhc} A.~Falkowski, J.~Ruderman, T.~Volansky, J.~Zupan, {\it in preparation}.

\bibitem{Acosta:2005ix}
D.~E.~Acosta {\it et al.}  [CDF Collaboration],
Phys.\ Rev.\  D {\bf 71} (2005) 112002
[arXiv:hep-ex/0505013].

\bibitem{Read:CLs}
A.L. Read, {\it ModiÞed frequentist analysis of search results (the CLs method)}, in
F. James, L. Lyons, and Y. Perrin (eds.), Workshop on ConÞdence Limits, CERN
Yellow Report 2000-005, available on cdsweb.cern.ch.

\bibitem{Junk:1999kv}
  T.~Junk,
  Nucl.\ Instrum.\ Meth.\  A {\bf 434}, 435 (1999)
  [arXiv:hep-ex/9902006].
  
  \end{thebibliography}
\end{document}